\documentclass[final, twocolumn]{elsarticle}
\usepackage[T1]{fontenc}

\usepackage{amsmath}
\usepackage{amssymb}
\usepackage{algorithm}
\usepackage{algorithmicx}
\usepackage{algpseudocode}
\usepackage{caption}
\PassOptionsToPackage{dvipsnames}{xcolor}

\usepackage{tikz}

\usetikzlibrary{shapes,arrows, decorations.text}

\tikzstyle{decision} = [diamond, draw, fill=blue!30, 
    text width=4.5em, text badly centered, inner sep=0pt]
\tikzstyle{block} = [rectangle, draw, fill=blue!20, 
    text width=12em, text centered, rounded corners, minimum height=4em, minimum width=10em]
\tikzstyle{start} = [rectangle, draw, fill=green!40, 
    text width=9em, text centered, minimum height=4em]
\tikzstyle{line} = [draw, -latex']
\tikzstyle{cloud} = [draw, ellipse,fill=red!20, minimum height=2em, 
    text width=9em, text centered]

\usepackage{dcolumn}
\usepackage{bm}
\usepackage[caption=false]{subfig}
\usepackage{hyperref}

\algrenewcommand\algorithmicindent{1.0em}

\definecolor{my_gray}{HTML}{7F7F7F}
\definecolor{my_yellow}{HTML}{CC6600}
\definecolor{my_green}{HTML}{27AF0B}
\definecolor{my_purple}{HTML}{93278F}

\newcommand{\eV}{\; \text{eV}}

\newcommand{\GeV}{\; \text{GeV}}

\newcommand{\gcm}{\; \text{g}/\text{cm}^2}

\newcommand{\m}{\; \text{m}}
\newcommand{\ns}{\; \text{ns}}

\newcommand{\Xmax}{X_{\text{max}}}
\newcommand{\DXmax}{\Delta X_{\text{max}}}
\newcommand{\vB}{\vec{v} \times \vec{B}}
\newcommand{\vvB}{\vec{v} \times \left( \vec{v} \times \vec{B} \right)}

\newcommand{\textMHz}{$\text{MHz}$}

\newcommand{\textXmax}{$X_{\text{max}}$}
\newcommand{\textDXmax}{$\Delta X_{\text{max}}$}
\newcommand{\textgcm}{$\text{g} / \text{cm}^2$}
\newcommand{\textcm}{$\text{cm}$}
\newcommand{\textm}{$\text{m}$}

\newcommand{\slice}{\text{slice}}
\newcommand{\ant}{\text{ant}}

\newcommand{\origin}{\text{origin}}
\newcommand{\synth}{\text{synth}}
\newcommand{\real}{\text{real}}

\journal{Astroparticle Physics}

\begin{document}

\begin{frontmatter}
    
\title{\texttt{SMIET}: Fast and accurate synthesis of radio pulses \\ from extensive air shower using simulated templates}

\author[iihe]{Mitja Desmet}
    \ead{mitja.desmet@vub.be}
\author[iihe]{Stijn Buitink}
\author[iihe,kit]{Tim Huege}
\author[kit]{Keito Watanabe}
    \ead{keito.watanabe@kit.edu}

\affiliation[iihe]{
    organization={Inter-University Institute For High Energies (IIHE), Vrije Universiteit Brussel (VUB)},
    addressline={Pleinlaan 2},
    city={Brussels},
    postcode={1050},
    country={Belgium}
}

\affiliation[kit]{
    organization={Institute for Astroparticle Physics, Karlsruhe Institute of Technology (KIT)},
    addressline={PO Box 3640},
    city={Karlsruhe},
    postcode={76021},
    country={Germany}
}

\begin{abstract}
    Interpreting the data from radio detectors for extensive air showers typically relies on Monte-Carlo based simulation codes, which, despite their accuracy are computationally expensive and present bottlenecks for analyses.
    To address this issue we have developed a novel forward model called template synthesis, which synthesises the radio emission from cosmic ray air showers in a matter of seconds.
    This hybrid approach uses a microscopically simulated, sliced shower (the origin) as an input. 
    It rescales the emission from each slice individually to synthesise the emission from a shower with different properties (the target).
    In this process it employs semi-analytical relations dependent on the shower age within the slice.
    We describe the connection between an antenna and a slice using the viewing angle and normalise the emission from every slice with respect to the air shower geometry using a set of scaling relations.
    In order to be able to change the arrival direction during synthesis, we adjust the phases based on the expected geometrical delays.

    We benchmark the method by comparing synthesised traces to CoREAS simulations over a wide frequency range of [30, 500] \textMHz . 
    We compare the signal amplitudes as well as the fluences.
    The synthesis quality is primarily influenced by the difference in \textXmax\ between the origin and target shower, \textDXmax.
    When $\DXmax \leq 100 \gcm$ , the scatter on the maximum amplitudes of the geomagnetic traces is at most 4\%.
    For the traces from the charge-excess component this scatter is smaller than 6\%.
    We also observe a bias with \textDXmax\ up to 5\% for both components, which appears to depend on the antenna position.
    Since the bias is symmetrical around $\DXmax = 0 \gcm$, we can use an interpolation approach to correct for it.

    We have implemented the template synthesis algorithm in an open-source Python package called \texttt{SMIET}, which includes all the necessary parameters to apply the method.
    Users only need to provide an origin shower.
    The package provides an implementation of the algorithm in \texttt{NumPy}, but also offers an alternative implementation in the \texttt{JAX} framework which makes the code fully differentiable.
    This package has been successfully tested with air showers with zenith angles up to 50\textdegree\ and can be used with any atmosphere, observation level and magnetic field.
    We demonstrate that the synthesis quality remains comparable to our main benchmarks across various scenarios and discuss potential use cases, including its use in machine-learning-based reconstructions.
\end{abstract}

\end{frontmatter}

\section{Introduction}

Studying cosmic rays at the highest energies is a challenge, primarily due to their rapidly declining flux with energy, which makes direct detection beyond $\sim 10^{14} \eV$ impractical. 
Rather we detect the large cascades of secondary particles produced when the cosmic rays interact in the atmosphere. 
These cascades, known as extensive air showers (EAS), span kilometres and primarily consist of photons, electrons, positrons and muons.
We can detect these particles using several different methods, each providing complementary information about the shower development (see for example subsection 2.4 in \cite{schroder_radio_2017}).

One important signal coming from EAS is the radio emission. 
Using this electromagnetic signal for detection offers several benefits \cite{Huege2016}. 
Modern digital techniques and computing hardware make it relatively straightforward to operate radio experiments. 
Additionally, radio equipment is cost-effective and easily deployed over the large areas required to collect enough statistics on the high-energy cosmic ray air showers. 
Furthermore, radio detectors have a very large duty cycle, meaning they can operate almost continuously.
Radio detection also acts as a complementary method to fluorescence imaging.
We can use both to make very accurate measurements of energy and the shower maximum, but the systematic uncertainties affecting them are completely different.
This explains why the Pierre Auger Observatory added a radio antenna to every station in their recent AugerPrime upgrade \cite{castellina_on_behalf_of_the_pierre_auger_collaboration_augerprime_2019}. 

The radio emission is mainly generated by the large number of electrons and positrons present in the shower. 
They interact with Earth's magnetic field and release Bremsstrahlung as a result.
Additionally there is radio emission associated to the electrons which are stripped off the atmospheric molecules as the shower passes through the atmosphere. 
This microscopic view of the emission from EAS can be reinterpreted in a macroscopic way \cite{scholten_macroscopic_2008}, offering a better intuition. 
From the macroscopic perspective, we identify two mechanisms. 
The dominant one in air is the geomagnetic contribution, which is the result of a time-varying current transverse to the shower propagation. 
The other mechanism is the charge-excess contribution. 
Originally described by Askaryan \cite{askaryan_coherent_1965} for dense media, it also provides a non-negligible contribution in air. 
This effect arises from a build-up of negative charge at the shower front, which also varies in time. 
Both contributions have a different polarisation angle, allowing us to decouple them \cite{Glaser2016}.

State-of-the-art simulation codes, such as CoREAS \cite{Heck1998, Huege2013} and ZHAireS \cite{sciutto_air_1999, alvarez-muniz_monte_2012}, are based on the microscopic approach. 
These simulate every particle in a Monte-Carlo based system, calculating the radio emission per particle and summing the result in every antenna. 
The resulting computation time thus scales linearly with the number of antennas as well as the number of simulated particles in the shower, which in turn correlates with the primary energy.
The required simulation runtimes already represent a bottleneck in our current analyses.
For the next generation of radio telescopes such at the Square Kilometer Array \cite{Dewdney2009}, whose number of antennas will be orders of magnitude higher than the current generation, this problem will only get worse.
While parallel computing can offer some help, it is clear we will need a more efficient simulation framework. 
One option would be to use the macroscopic description, which can be treated analytically. 
Such simulation codes do exist, such as MGMR3D \cite{Scholten_analytic_2018}, but while they can be used to reconstruct air shower parameters, they produces biases that need to be corrected for \cite{mitra_high_2021}.

In our previous work, we introduced a novel approach to simulate the radio emission from EAS called template synthesis \cite{desmet_proof_2024}.
It is a hybrid approach, where we use elements from both the micro- as well as the macroscopic interpretations.
This allows us to benefit from the precision of the microscopic simulations, while still achieving much faster runtimes.
In \cite{desmet_proof_2024} we showed that we can produce EAS-induced radio pulses on the timescales of seconds, compared to the hours or even days required using microscopic simulations, while achieving an accuracy comparable to the inherent shower-to-shower fluctuations.
However, the method was constrained to a single air shower geometry.
Building on this proof of concept, we now present a generalised version of the template synthesis method.
This version is applicable to showers with any geometry, up to at least 50\textdegree\ zenith angle.
The amplitudes of the synthesised pulses match those from microscopic simulations within 9\% .
The template generation (which needs to be done only once) takes a minute or two, while the actual synthesis only takes a few seconds.
We also include an explicit treatment of the phases based on the arrival time calculated from the geometry.
This allows us to changes the zenith angle up to 4\textdegree\ during synthesis, while incurring at most a 5\% error, opening the possibility of arrival direction reconstruction studies.
The algorithm is now implemented in the \texttt{SMIET} (where the ``ie'' is pronounced as the ``ee'' from ``to meet'') software, which is publicly available as a Python package \cite{smiet_cr_synthesis}.

We view template synthesis as complementary to the Radio Morphing technique \cite{Zilles2020}.
While we target the more vertical showers with template synthesis, Radio Morphing becomes valid from 60\textdegree\ zenith angle onwards.
Both methods rescale the emission from a microscopically simulated input shower to synthesise pulses.
One important difference however, is that template synthesis allows a user to choose the longitudinal profile of the target shower.
In Radio Morphing the shower can only be shifted around.

This paper is organised in the following sections.
In the next Section \ref{sec:method} we introduce the general template synthesis method.
We cover our description of the geometry and the corrections we apply because of it.
We also discuss the spectral functions, which lie at the heart of the template synthesis approach.  %
In Section \ref{sec:workflow} we present the template synthesis software package and its workflow from a user perspective.
Then in Section \ref{sec:benchmarking} we test our method on air showers up to 50\textdegree\ zenith angle.
This will allow us to define some practical guidelines on how to use template synthesis.
The benchmarks are performed with the same showers from which the spectral functions were extracted however, so in Section \ref{sec:universality} we proceed to test template synthesis under different conditions.
We show that over a wide range of scenarios the performance of our method is the same.
The situations we cover also give a taste for the use cases that can be accommodated with template synthesis.
We cover these shortly in Section \ref{sec:discussion}, where we also discuss the differentiability of template synthesis as well as some implications our results have for radio emission from air showers in general.
We conclude in Section \ref{sec:conclusion}.

\section{The generalised template synthesis method} \label{sec:method}

The premise of template synthesis is to start from a single microscopically simulated shower, called the origin, and use scaling relations to synthesise the emission for a shower with different parameters.
This concept is introduced in subsection \ref{subsec:fundamentals}.
In previous work \cite{desmet_proof_2024} we presented a proof of concept for this method, but it had several limitations, which we review in subsection \ref{subsec:recap}.

In this paper we describe the generalised template synthesis method. 
Our goal was to update the procedure from \cite{desmet_proof_2024} to make it applicable to any air shower geometry. 
To this end, we have changed the geometrical description in subsection \ref{subsec:geometry} to relate the antenna positions to the EAS using the viewing angle. 

Next, in subsection \ref{subsec:scaling_relations}, we introduce a new set of scaling relations which remove the dependency of the radio emission on the EAS geometry. 
This will allow us in subsection \ref{subsec:spectral_functions} to extract the generalised \textit{spectral functions}. 
These can be applied to all geometries of interest. 
We also detail a scheme to interpolate between the spectral functions in subsection \ref{subsec:interpolation}.

One major improvement we made compared to \cite{desmet_proof_2024} is an explicit treatment of the phases, which we discuss in subsection \ref{subsec:phases}.
Finally, in subsections \ref{subsec:template} and \ref{subsec:synthesis} we discuss how to apply template synthesis to obtain a synthesised radio signal.

\subsection{Fundamentals of template synthesis} \label{subsec:fundamentals}

At the centre of the template synthesis method is the concept of sliced simulations.
These are simulations where we divide the atmosphere into slices of constant atmospheric depth.
In CoREAS, with standard functionality, we can record the emission from every slice $\vec{E}_{\slice}$ individually.
The complete signal in an antenna at distance $\vec{r}$ can then be reconstructed by summing the contributions from all slices,
\begin{equation}
    \vec{E} ( \vec{r}, t ) = \sum_{X} \vec{E}_{\text{slice}} ( X, \vec{r}, t ) \; .
\end{equation}
Here we differentiate the slices by their atmospheric depth $X$.
Besides the emission, we also record the number of electrons and positrons $N$ in each slice.
The resulting function $N(X)$ we refer to as the longitudinal profile.

The goal of the template synthesis method is to rescale the emission from one microscopically simulated shower to reproduce the emission of a shower with a different longitudinal profile and primary particle.
The sliced microscopic simulation that serves as input to the method, we call the \textit{origin} shower.
The \textit{target} shower on the other hand, is the one for which we want to synthesise the emission.

The template synthesis procedure is applied to each slice separately.
First, it accounts for the number of electrons and positrons in the slice, $N_{\slice}$.
This rescaling largely removes the dependency of the emission amplitude on the primary energy of the cosmic ray, since the number of particles scales approximately linearly with the energy.

It turns out however, that a correction to the pulse shape is also necessary. 
In \cite{desmet_proof_2024} we observed a correlation between the \textXmax\ of the shower and the shape of the emission amplitude frequency spectrum within each slice.
We interpreted this as a dependency of the pulse shape on the shower age in the slice.

This correction to the pulse shape is applied to the amplitude frequency spectra using the spectral functions, which are discussed in more detail in subsection \ref{subsec:spectral_functions}.
These give the expected shape of the amplitude spectra, given the slice and shower age.
The spectral functions are semi-analytical expressions that we extracted from a large set of Monte-Carlo simulations and which can then be universally used for template synthesis.

\begin{figure}
    \centering
    \includegraphics[width=\linewidth]{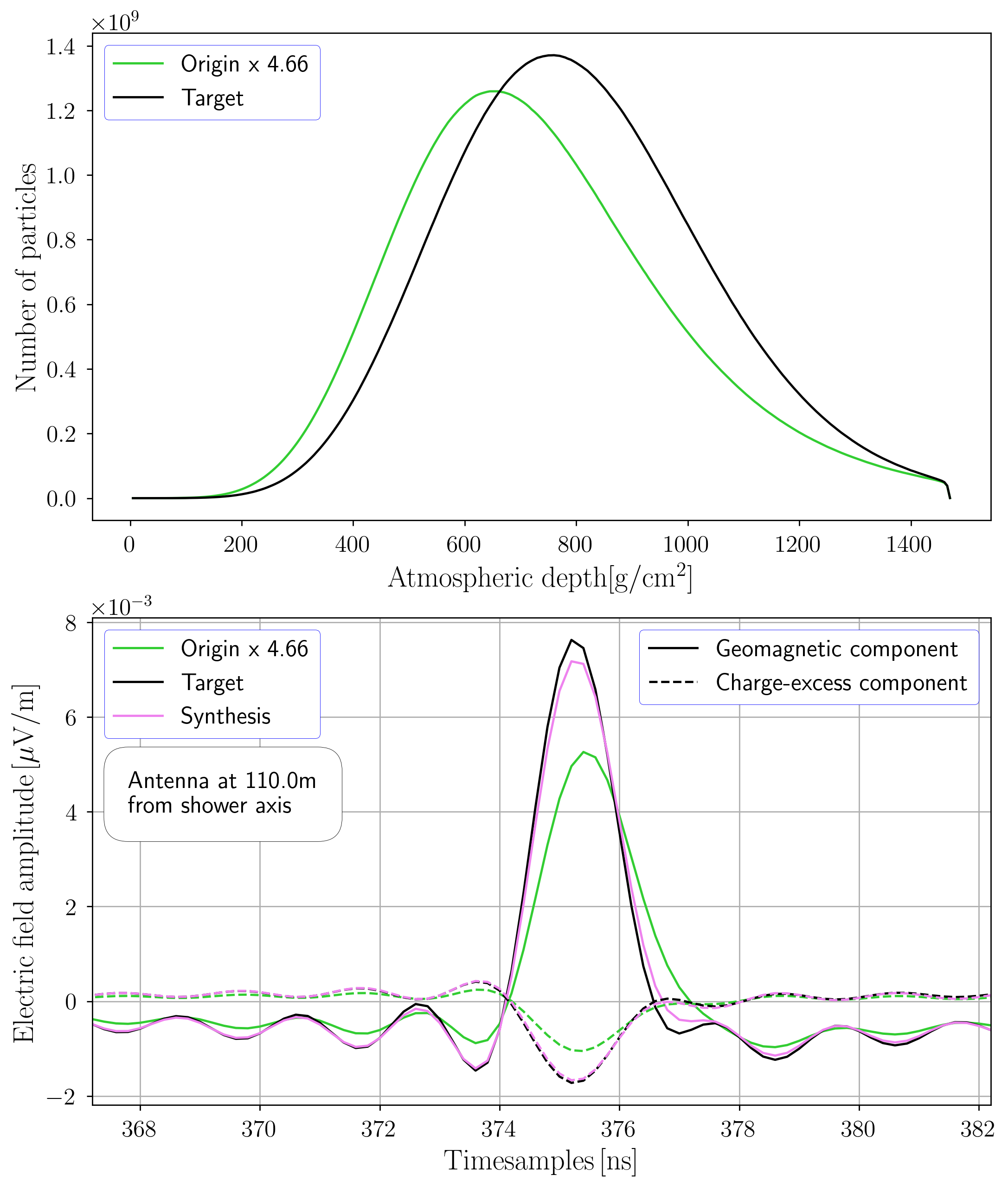}
    \caption{
        An example of applying template synthesis for an antenna at 110 \textm\ from the shower axis (in the shower plane).
        Here the origin is initiated by a $4.69 \times 10^8 \GeV$ iron primary, while the target shower has a proton primary with an energy of $2.19 \times 10^9 \GeV$.
        Both have a 45\textdegree\ zenith angle and an azimuthal angle of 0\textdegree\ (in the CORSIKA coordinate system).
        Their \textXmax\ values differ by 102 \textgcm .
        In this case we have simulated both showers with CoREAS, in order to compare the result of template synthesis to that of CoREAS.
        On top we show the longitudinal profiles.
        The origin profile (in green) has been multiplied with the ratio of the primary energy of both showers for clarity.
        The bottom plot shows the radio signals in the antenna, which have been filtered using a [30, 500] \textMHz\ bandpass filter, where we decoupled the emission into the geomagnetic and charge-excess components.
        Again in green we have the signal from the origin shower, which serves as the input to the method, multiplied by the ratio of the primary energies.
        In black we show the trace of the target shower as simulated by CoREAS, which is the reference we try to approach using template synthesis.
        The result of the latter is shown in magenta.
        We can see that template synthesis is capable of not only adjusting the amplitude of the signal to match that of the CoREAS one, but also alter the shape of the pulse.
    }
    \label{fig:example_synthesis}
\end{figure}

Figure \ref{fig:example_synthesis} shows an example of applying template synthesis.
We can see that the synthesised signal aligns closely with the result from a full Monte-Carlo simulation.
We can also appreciate that template synthesis not only adjusts the amplitude of the input signal, but alters the pulse shape as well.
A rigorous benchmark \cite{desmet_proof_2024} showed that in our proof-of-principle study, the method achieved an accuracy comparable to the intrinsic shower-to-shower fluctuations, which are around the 4\% level.
This is the amount of scatter on the maximum amplitude we found when comparing air showers with very similar longitudinal profiles in CoREAS.
Since we use CoREAS simulation as inputs, we expect template synthesis to be limited by these fluctuations.

\subsection{Limitations of the previous approach} \label{subsec:recap}

In \cite{desmet_proof_2024} we were able to change the longitudinal profile of the showers (and through it also the primary particle type and energy).
The arrival direction was however fixed, i.e. all showers came from zenith.
We could not vary the direction, because the spectral functions were unique to the geometry of the air showers used to extract them.
This meant that for every geometry of interest, one would have to recalculate them.
This entailed a large computational cost, which made it impractical for actual applications.

The fundamental limitation in \cite{desmet_proof_2024} was that we labelled each slice by the atmospheric depth at the bottom of that slice and used that to describe the relation between the slices and the antennas.
Because the atmospheric depth of a given slice changes when altering the air shower geometry, this description does not allow us to alter the latter.

Furthermore, the phase frequency spectra were left untouched.
This was done with the assumption that the information in the phase spectra is mostly geometrical.
And since we were working with a fixed geometry, there was no need to change it.

Hence, when generalising the method, we adapted these two aspects.
First we changed the description of how to relate an antenna to a slice.
This is described in the next subsection.
Then in subsection \ref{subsec:phases} we introduce an explicit treatment of the phases.

\subsection{A geometrical setup based on viewing angle} \label{subsec:geometry}

In order to characterise the dependency on the shower age, we will change our way of labelling the slices.
From now on we will use $\DXmax^{\slice}$, the distance of the bottom of the slice to the air shower maximum \textXmax\ in \textgcm .
This quantity acts as a proxy for the age of the shower in a slice.
This means that when comparing the emission from two different showers, we will compare the slices with the same distance to \textXmax\ (within the slicing precision).
We note that this change from establishing spectral functions for each given pair of ($X_{\slice}, \Xmax$) as in \cite{desmet_proof_2024}, to now only spectral functions for every $\DXmax^{\slice}$ fundamentally signifies that the pulse shape is indeed driven by the shower age only, a hypothesis we had already phrased in \cite{desmet_proof_2024} but now confirmed explicitly, thus simplifying and generalising the template synthesis approach.

A challenge we face when trying to compare the radio emission from two EAS with different geometries is the question of which antenna positions would receive the same signal, if all other parameters were kept fixed between the two cases?
If we think about this in the context of sliced simulations, it makes sense to consider the opening angle of the receiving antenna with respect to a slice.
Here the opening angle is the angle between the shower axis and the straight line connecting the (bottom of the) slice and the receiver.
Supposing a slice with a given $\DXmax^{\slice}$ always emits in the same way, we could conclude antennas which observe the slice under the same opening angle would be equivalent.

However, the emission pattern of a slice will depend on the EAS geometry.
As for different zenith angles, slices with the same slant depth will have different altitudes.
Along with the height, the density and refractive index in the slice will change.
The latter has a direct implication on the beaming of the radiation.
Due to the relativistic nature of the emitting particles, the emission is strongly beamed forward along the Cherenkov angle.
The Cherenkov angle $\theta^C$ is related to the refractive index $n$ as
\begin{equation}
    \theta^C_{\slice} = \arccos \left( \frac{1}{n_{\slice}} \right) \; ,
\end{equation}
where the subscripts refer to the values in a slice.
We will refer to these parameters as the \textit{local} values, in the sense that they are local to a single slice.

Hence, the equivalent antenna will not be the one which sits at the same opening angle.
Instead we define the \textbf{normalised viewing angle} as the opening angle divided by local Cherenkov angle.
We will use this unitless quantity to define an antenna's position.
This is analogous to the transformation applied in the Radio Morphing approach \cite{Zilles2020}, except that in our case one antenna has a different viewing angle with respect to each slice.

\begin{figure}
    \centering
    \includegraphics[width=\linewidth]{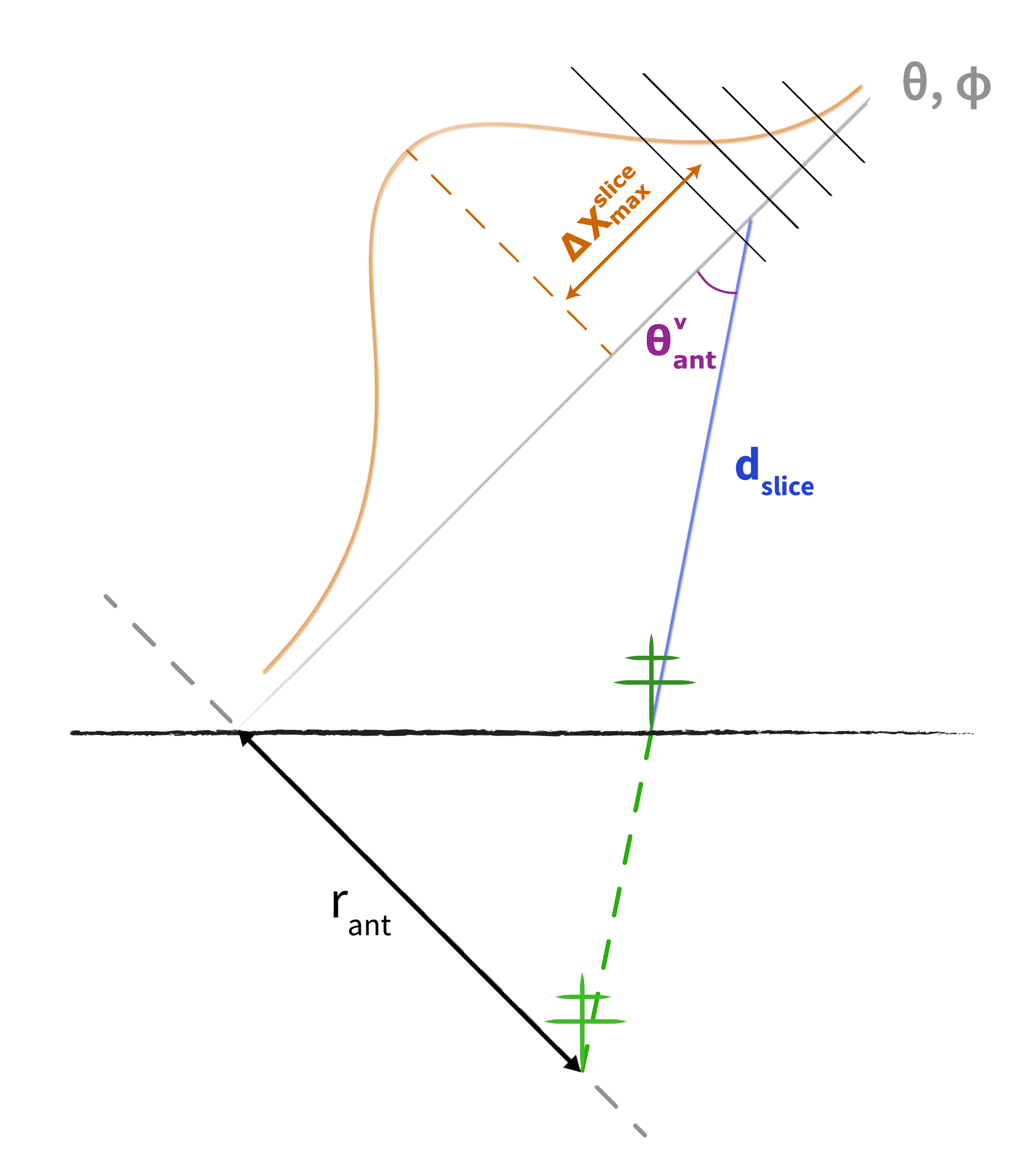}
    \caption{
        The shower axis is defined by the \textcolor{my_gray}{zenith angle} $\theta$ and \textcolor{my_gray}{azimuth angle} $\phi$ of the shower. 
        The angle between the shower axis and Earth's magnetic field vector is the geomagnetic angle $\alpha_{\text{GEO}}$.
        The slices are taken perpendicular to the shower axis.
        A slice is characterised by its \textcolor{my_yellow}{distance to shower maximum} $\DXmax^{\slice}$ (in \textgcm ).
        For every slice we can retrieve or calculate its properties, like the number of emitters $N_{\slice}$, the local air density $\rho_{\slice}$ and the local Cherenkov angle $\theta^C_{\slice}$. 
        The relation between a receiver and a slice is encoded by the \textcolor{my_purple}{viewing angle} $\theta^v_{\ant}$, which is the opening angle in units of the local Cherenkov angle of the slice, together with the \textcolor{Blue}{geometrical distance to the antenna} $d_{\slice}$.
        Perpendicular to the shower axis, we define the shower plane.
        We can project the antenna into the shower plane along the line of sight from the slice to the antenna.
        We use the theoretical polarisation patterns of the geomagnetic and charge-excess components in this plane to decouple these two contributions.
    }
    \label{fig:geometry}
\end{figure}

As such we arrive at the setup shown in Figure \ref{fig:geometry}.
Since everything is expressed with respect to the EAS properties, we can easily compare between different showers.
For this we will compare the emission from slices with same distance to the shower maximum, in antennas which are positioned under the same viewing angle, respectively.

\subsection{Rescaling the emission with respect to the EAS geometry} \label{subsec:scaling_relations}

\begin{figure*}
    \centering
    \subfloat[The amplitude frequency spectra from showers with different zenith angles without any scaling.]{
        \includegraphics[width=0.8\textwidth, trim={0 0 0 2cm}, clip]{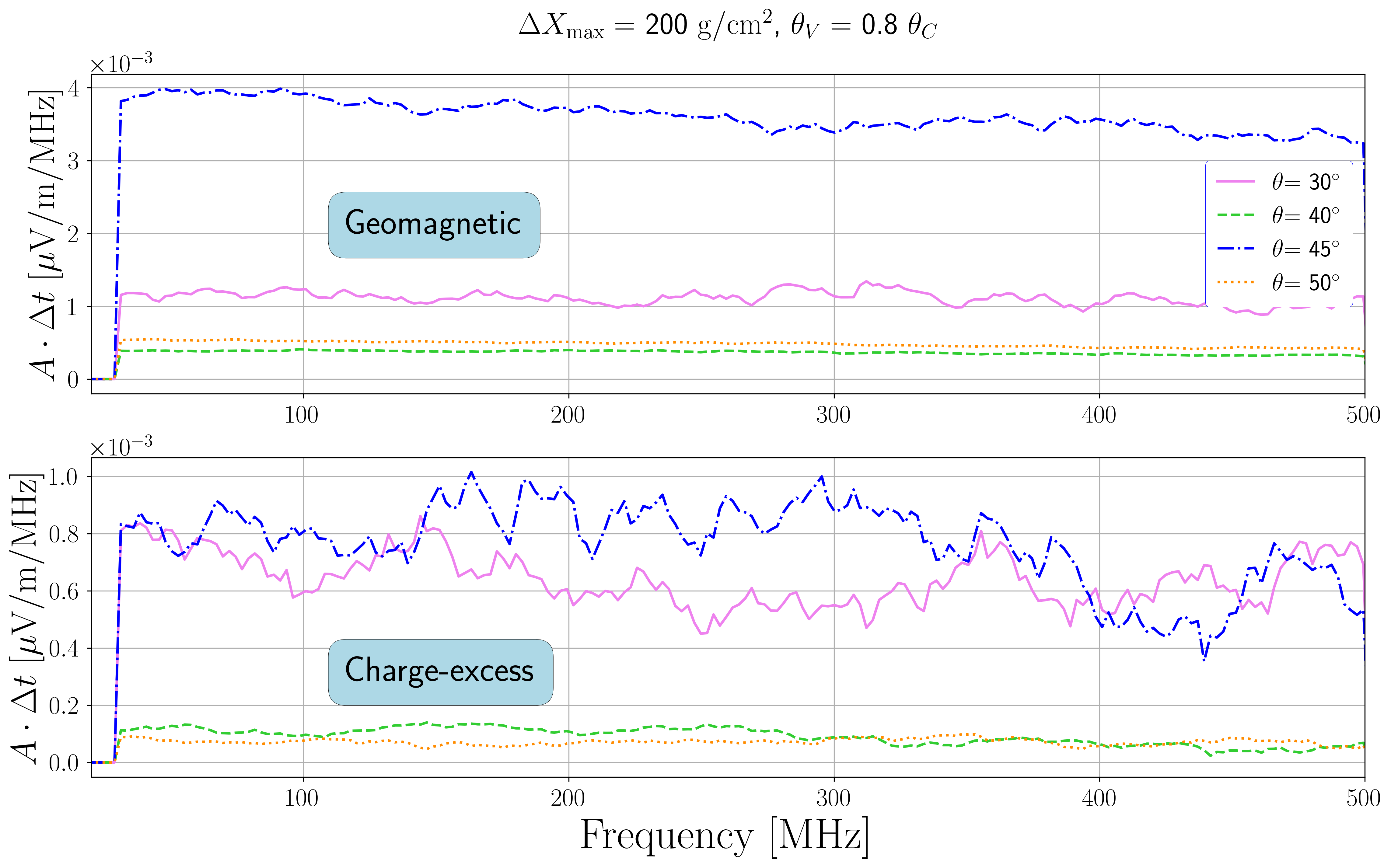}
        \label{fig:spectra_unscaled}
    } \\
    \subfloat[The same spectra as shown in (a) after applying the scaling relations mentioned in the main text.]{
        \includegraphics[width=0.8\textwidth, trim={0 0 0 2cm}, clip]{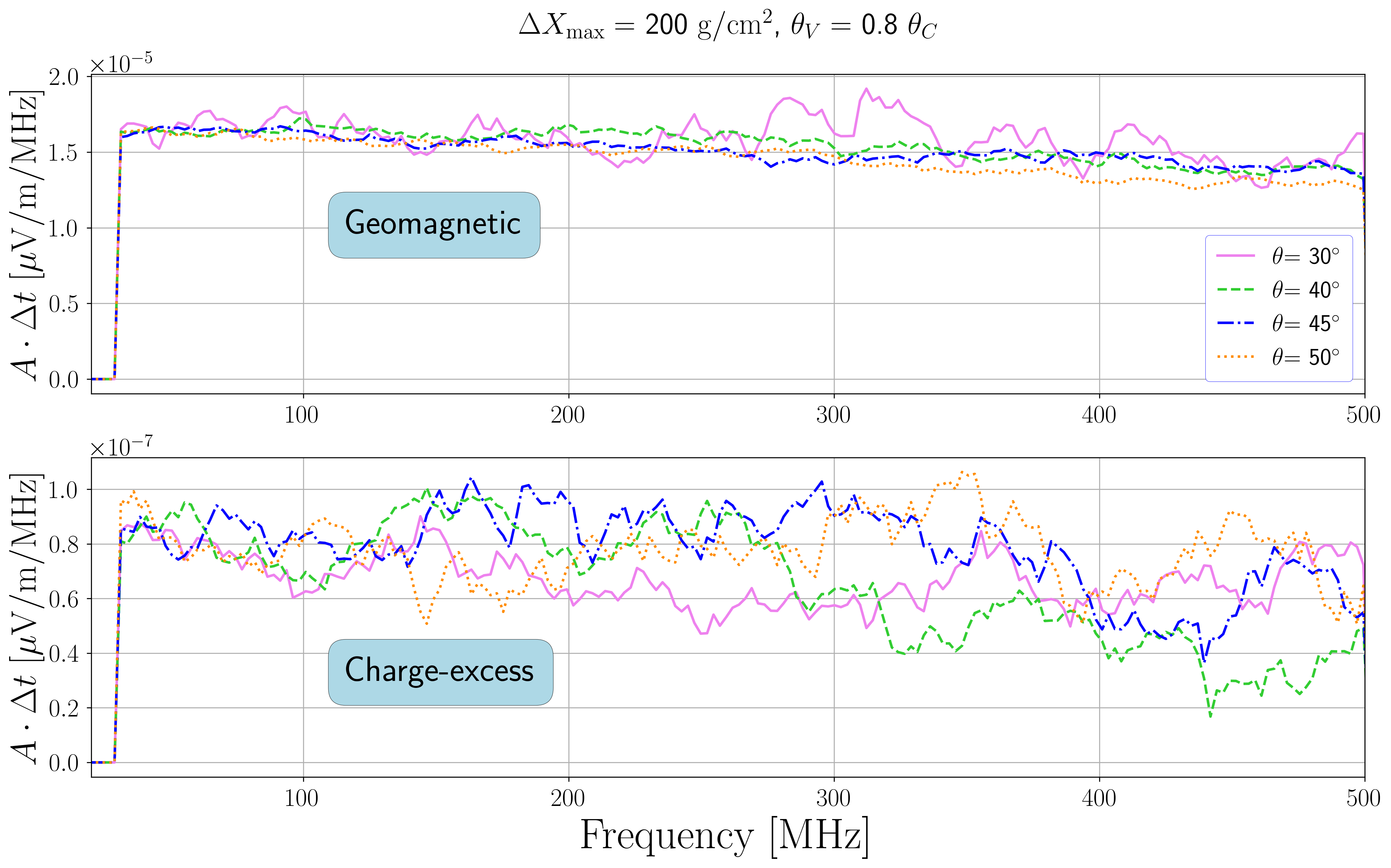}
        \label{fig:spectra}
    }
    \caption{
        Spectra from four different air showers, each with a different zenith angle (while keeping the azimuth angle fixed), simulated using the settings mentioned in subsection \ref{subsec:extracting}.
        From each of them, we look at the geomagnetic (GEO, top) and charge-excess (CE, bottom) amplitude frequency spectra coming from the slice with $\DXmax^{\slice} = -200 \gcm$, in an antenna with $\theta_{\ant}^{v} = 0.8$. 
        In (a) we see that without correcting for the geometry, the spectra have a seemingly random ordering. 
        Once we apply the scalings mentioned in the main text in (b) however, all spectra overlap and can be described by a single function.
    }
\end{figure*}

With the normalised viewing angle we can already find the antennas in which the amplitude spectrum from a slice where a shower has some given age, has the same shape.
However, it does not yet fix the normalisation of the spectrum.
Indeed, when we compare the amplitude frequency spectra from slices with the same $\DXmax^{\slice}$ in antennas with the same viewing angle in Figure \ref{fig:spectra_unscaled}, we see they differ strongly.
We can already expect the physical distance from the slice to the receiver to play a role.
But there are also other effects, which we will need to explicitly account for.

The first one of these, which we already included in \cite{desmet_proof_2024}, is the number of emitters $N_{\slice}$.
Next, we have that the geomagnetic component scales with the sine of the geomagnetic angle, which is the angle between the shower axis and the Earth's magnetic field vector \cite{pierre_auger_measurement_2016}.

In \cite{ammerman-yebra_density_2023} the authors investigated the dependency of the amplitude of the radio emission on the air density and magnetic field strength. 
They found relations which described how the amplitude scaled with these parameters. 
However, they looked at the emission coming from entire showers.
In our testing we found that only two of the scaling relations were relevant for our sliced emission.

The first one is the scaling of the geomagnetic amplitude with the inverse of the air density.
As the density drops, the Lorentz force which is responsible for the transverse current producing the geomagnetic component acts for longer periods of time.
Since the average time between the particle interactions grows as the density lowers, the drift velocity increases and the current grows \cite{scholten_macroscopic_2008}.
This scaling relation holds for densities down to $600 \, \text{g} \, \text{m}^{-3}$, which is about half the sea level density.
For lower densities this starts to break down, because the lateral extent of the shower becomes so large that the coherency of the particles is lost.

We do note that in \cite{ammerman-yebra_density_2023} the authors also shifted the frequencies with the increase in density to line up certain spectral features.
However, we did not observe this shift in our tests, when comparing the amplitude spectra from different showers in one slice.
We surmise the reason for this might be that the aforementioned spectral features are caused by the interference of particles from the whole shower.
In this case, these features would not appear on the level of individual slices.
Hence we only apply a density correction to the amplitude of the geomagnetic component and not the frequencies associated to the spectrum.

For the charge-excess emission the authors of \cite{ammerman-yebra_density_2023} observed a dependency of the amplitude on the sine of the local Cherenkov angle.
This was interpreted as a projection factor of the particle velocity into the plane perpendicular to the observing direction, since their observer was always at the Cherenkov angle.
We found however that in our setup this scaling could be applied to all antennas.\footnote{We compared using the opening angle of the antenna instead of the Cherenkov angle of the slice for the plots like the one shown in Figure \ref{fig:spectra}, and observed a very similar relative scatter. Because the Cherenkov angle is easier to work with, we opted to use that.}
This is probably because in our setup we always compare antennas under the same viewing angle, which are expressed as multiples of local Cherenkov angle. 

Overall we arrive at the following set of scaling relations.
For the geomagnetic emission, we rescale the emission by
\begin{itemize}
    \item the inverse of the distance from slice to antenna, $d^{-1}_{\slice}$, 
    \item the number of emitting particles in the slice, $N_{\slice}$,
    \item the sine of the geomagnetic angle, $\sin\left( \alpha_{\text{GEO}} \right)$ and 
    \item the inverse of the air density, $\rho^{-1}_{\slice}$ .
\end{itemize}
We rescale the charge-excess emission on the other hand by
\begin{itemize}
    \item the inverse of the distance from slice to antenna, $d^{-1}_{\slice}$, 
    \item the number of emitting particles in the slice, $N_{\slice}$ and
    \item the sine of the local Cherenkov angle, $\sin\left( \theta^C_{\slice} \right)$ .  
\end{itemize}
Applying these to the amplitude frequency spectra in Figure \ref{fig:spectra_unscaled}, of which the result is shown in Figure \ref{fig:spectra}, we indeed see that the spectra overlap.
This means we have successfully removed the dependency on the geometry when comparing the emission from slices with the same $\DXmax^{\slice}$ in antennas under the same viewing angle.

\subsection{Extracting the spectral functions from microscopic simulations} \label{subsec:spectral_functions}

Given that we now can rescale the spectra to be independent of the air shower geometry, we can proceed to fit their dependency on shower age.
In order to this we start from a large set of sliced Monte-Carlo simulations.
This set should preferably contain a mixture of primary particle types and energies, and also have different zenith angles.
Varying the azimuthal angle is also possible, but this only induces a change in the geomagnetic angle and a rotation of the antennas on the ground.
Both these effects are well understood.

We now construct the spectral functions.
These capture the trends in the microscopically simulated set.
Template synthesis then uses the information from these spectral functions to synthesise the pulses.
This is why it is important that the simulation set used to extract the functions covers the parameter range of interest.

The procedure we describe here is applied with a single, fixed normalised viewing angle $\theta^v_{\ant}$.
This means that in CoREAS, we want to configure one observer for each slice in such a way that it always observes the slice under the chosen viewing angle.
Alternatively, we can simulate an array of observers for each slice and then use interpolation to obtain the antenna at the viewing angle we want.
Of course in practice, such as in Section \ref{sec:benchmarking}, we will want to have the spectral functions for multiple viewing angles.
For this we simply apply the procedure repeatedly, for each angle we want.

The first step is to fit the amplitude frequency spectrum from each slice.
The fit functions we use are similar to the ones presented in \cite{desmet_proof_2024}, with only the extra scalings from above incorporated.
We thus have the following two parametrisations for the geomagnetic (GEO) and the charge-excess (CE) component respectively:
\begin{align}
    \nonumber \tilde{A}_{\text{geo}} &(f) 
    = 
    \left( 
        a_{\text{geo}} \cdot \frac{N_{\text{slice}} \cdot \sin(\alpha_{\text{GEO}})}{d_{\text{slice}} \cdot \rho_{\text{slice}}} 
    \right) \\
    & \; \times \exp \left( b_{\text{geo}} \cdot (f - f_0) + c_{\text{geo}} \cdot (f - f_0)^2 \right) \; , \label{eq:a_geo}\\[2ex]
    \nonumber \tilde{A}_{\text{ce}} &(f) 
    = \left( a_{\text{ce}} \cdot \frac{N_{\text{slice}} \cdot \sin(\theta^{\text{C}}_{\slice})}{d_{\text{slice}}} \right) \\& \, \times \exp \left( b_{\text{ce}} \cdot (f - f_0) \right) \label{eq:a_ce} .
\end{align}
Here $f$ denotes frequency.
We perform these fits over some fixed range in frequency, as the spectra exhibit quite distinct features in different bands.
The ``central frequency'' $f_0$ is subtracted to ensure the values of the constants before the exponential are determined in a physical region of interest\footnote{Typical experiments do not consider frequencies below 30 \textMHz\ because of noise from the ionosphere. Also, the amplitude is expected to be zero at 0 \textMHz\ as there is no static charge.}.
Usually it is best to choose its value between 50 and 100 \textMHz , as most spectra have a significant signal in that region.
But it should also lie within the frequency band chosen previously.

We refer to $a$, $b$ and $c$ as the \textbf{spectral parameters}.
They depend on the slice they were extracted from, as well as the antenna which is characterised by its viewing angle $\theta^v_{\ant}$,
\begin{align*}
    & a(\theta^v_{\ant}, \DXmax^{\slice}) \; , \\
    & b(\theta^v_{\ant}, \DXmax^{\slice}) \; , \\
    & c(\theta^v_{\ant}, \DXmax^{\slice}) \; .
\end{align*}

\begin{figure*}
    \centering
    \includegraphics[width=\linewidth, trim={0 0 0 1cm}, clip]{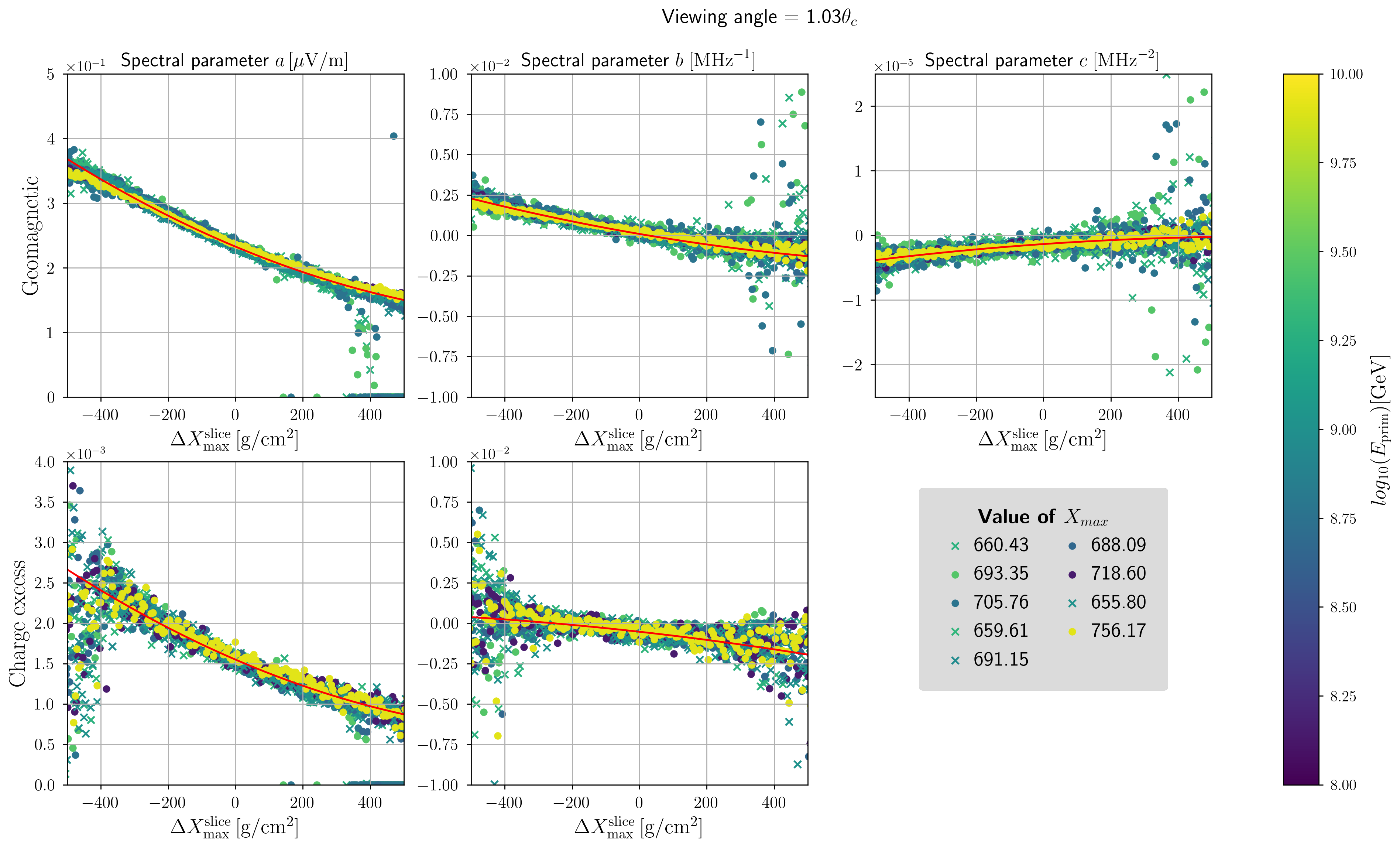}
    \caption{
        Here we show an example of how the spectral parameters depend on the $\DXmax^{\slice}$ of the slice they were obtained from.
        As a reminder, negative $\DXmax^{\slice}$ refer to slices earlier than the shower maximum, whereas positive values refer to slice after it.
        We show the values obtained from 9 different showers, three with a zenith angle of 30\textdegree , three with a zenith angle of 40\textdegree\ and another three with a zenith angle of 50\textdegree .
        The colour refers to the primary energy and the marker style to primary particle type, with dots being proton induced showers and crosses iron induced ones.
        All points with the same colour are spectral parameters obtained from the same shower, but in different slices.
        In each slice we considered an antenna which was placed at 1.03 times the local Cherenkov angle.
        These values were obtained by fitting showers from the simulation described in Section \ref{sec:benchmarking}, over a frequency range of [30, 500] \textMHz .
        We see that all points fall onto the one parabola, indicated by a red line, irrespective of the primary energy or type, as well as the \textXmax\ of the shower.
        This red line is what we call the spectral function.
    }
    \label{fig:spectral_fits}
\end{figure*}

The second step is to characterise the dependency of each spectral parameter to the $\DXmax^{\slice}$ of the slice they came from.
In Figure \ref{fig:spectral_fits} we show the evolution of the spectral parameters, for a normalised viewing angle of $1.03$.
Based on the observed trend, we fit a parabola to these points.
To better deal with the large scatter from slices with few particles, we bin the data points in $\DXmax^{\slice}$ to obtain a mean value and its standard deviation.
If it should happen that a bin contains less than two data points, which is possible for very early or late slices where only a few showers have particles, it is not considered for the fit.

The result of the fit is the \textbf{spectral function} of a spectral parameter.
The spectral functions encode the dependency of the spectral parameters on the $\DXmax^{\slice}$ of the slices, for the viewing angle under consideration,
\begin{align*}
    & a(\theta^v_{\ant}, \DXmax^{\slice}) = p_0^a + p_1^a \cdot \DXmax^{\slice} + p_2^a \cdot (\DXmax^{\slice})^2 \\
    & b(\theta^v_{\ant}, \DXmax^{\slice}) = p_0^b + p_1^b \cdot \DXmax^{\slice} + p_2^b \cdot (\DXmax^{\slice})^2 \\
    & c(\theta^v_{\ant}, \DXmax^{\slice}) = p_0^c + p_1^c \cdot \DXmax^{\slice} + p_2^c \cdot (\DXmax^{\slice})^2 \; .
\end{align*}
We can now calculate the expected spectral parameters for any slice, in the antenna with the corresponding viewing angle, using only the $p$-parameters.
And from the spectral parameters we can then calculate expected amplitude frequency spectrum.
In total we have 15 $p$-parameters for one viewing angle.
These are 9 for the GEO components and 6 for the CE, as the latter does not have the $c$ spectral parameter.
The $p$-parameters essentially hold all the microscopic information we need to apply template synthesis.

\subsection{Interpolation of the spectral functions} \label{subsec:interpolation}

Before we can use these spectral functions in practice, we need to address one subtle issue.
Like mentioned in the previous section, the spectral functions are for a fixed viewing angle.
When we process any shower, it is unlikely that it will have its antennas placed at exactly the viewing angles we have stored.\footnote{Also, an antenna fixed on the ground will have different viewing angles for every slice.}
Hence we need to interpolate the spectral functions.

This will influence the choice of which normalised viewing angles to process and store, in a practical implementation of template synthesis such as in Section \ref{sec:workflow}.
We know that at the Cherenkov angle, the amplitude spectra are quite flat.
On the other hand, as we move away from the angle the spectra become steeper.
The shapes of the spectra change rapidly at first, but the farther from the Cherenkov angle we are, the slower the rate of change.
Therefore, we suggest to construct the spectral functions for normalised viewing angles ranging from $0.1$ to $2.0$, with a denser sampling around $1.0$ (which is exactly at the local Cherenkov angle).
We can still extrapolate outside of this range by using the spectral function for the closest viewing angle, as from the edges onwards, the spectra are quite similar in shape and only differ in normalisation.
Furthermore, there will be no significant signal far away from the Cherenkov angle.

In order to interpolate against the viewing angle, we followed a similar approach to the Fourier-based signal interpolation presented in \cite{corstanje_high-precision_2023}.
There, the amplitude frequency spectra are interpolated per frequency bin.
We apply this technique in each slice individually.
Per slice we evaluate the spectral functions, yielding a spectrum per viewing angle. 
To interpolate between viewing angles, we employ a linear interpolation scheme per frequency bin.
This is different to the cubic splines used in the Fourier interpolation.
We opted for the linear approach because of the rapid changes close to the Cherenkov angle and the cubic splines might introduce wiggles that are non-physical.\footnote{We did try the cubic splines approach and only noticed some issues with synthesised pulses in antennas very close to the shower axis.}
When extrapolating outside of the provided range of normalised viewing angles, we use the values at the edges.
For example, if we have the spectral functions up to a normalised angle of $2.0$ and request a viewing angle of $3.0$, we return the values for the case of $\theta_{\ant}^v = 2.0$.

The interpolation of the amplitude spectra will not be mentioned explicitly in the following, we will simply assume we can obtain the normalised spectra for any viewing angle we need.

\subsection{Accounting for the phases} \label{subsec:phases}

One aspect we have not discussed so far is the treatment of the phase spectra.
In previous publications, we took the phases from the origin shower and did not alter them when synthesising a target shower.
The reasoning was that the information contained in the phases is mostly geometrical and since we were always using an origin shower with the same geometry as the target we wished to synthesise, it was not necessary to modify it.
However, in order to use template synthesis for event reconstruction, it should be capable of varying the geometry at least within the uncertainty of the arrival direction estimated by classical methods, i.e., within a maximum of few degrees.
For example, LOFAR has an arrival direction estimate which has an uncertainty of $\sim 1$\textdegree\ from the LORA particle detector.
Using this information we could generate an appropriate origin shower, i.e. with the arrival direction estimated from the particle detector, which then serves as the starting point for direction reconstruction.
We explore this use case partially in subsection \ref{subsec:diff_geometry}.

\begin{figure*}
    \centering
    \subfloat[Phase spectra in an antenna at 140\textm\ from the shower axis, from multiple showers with different \textXmax .]{
        \includegraphics[width=0.6\linewidth, trim={0 0 6.0cm 2.7cm}, clip]{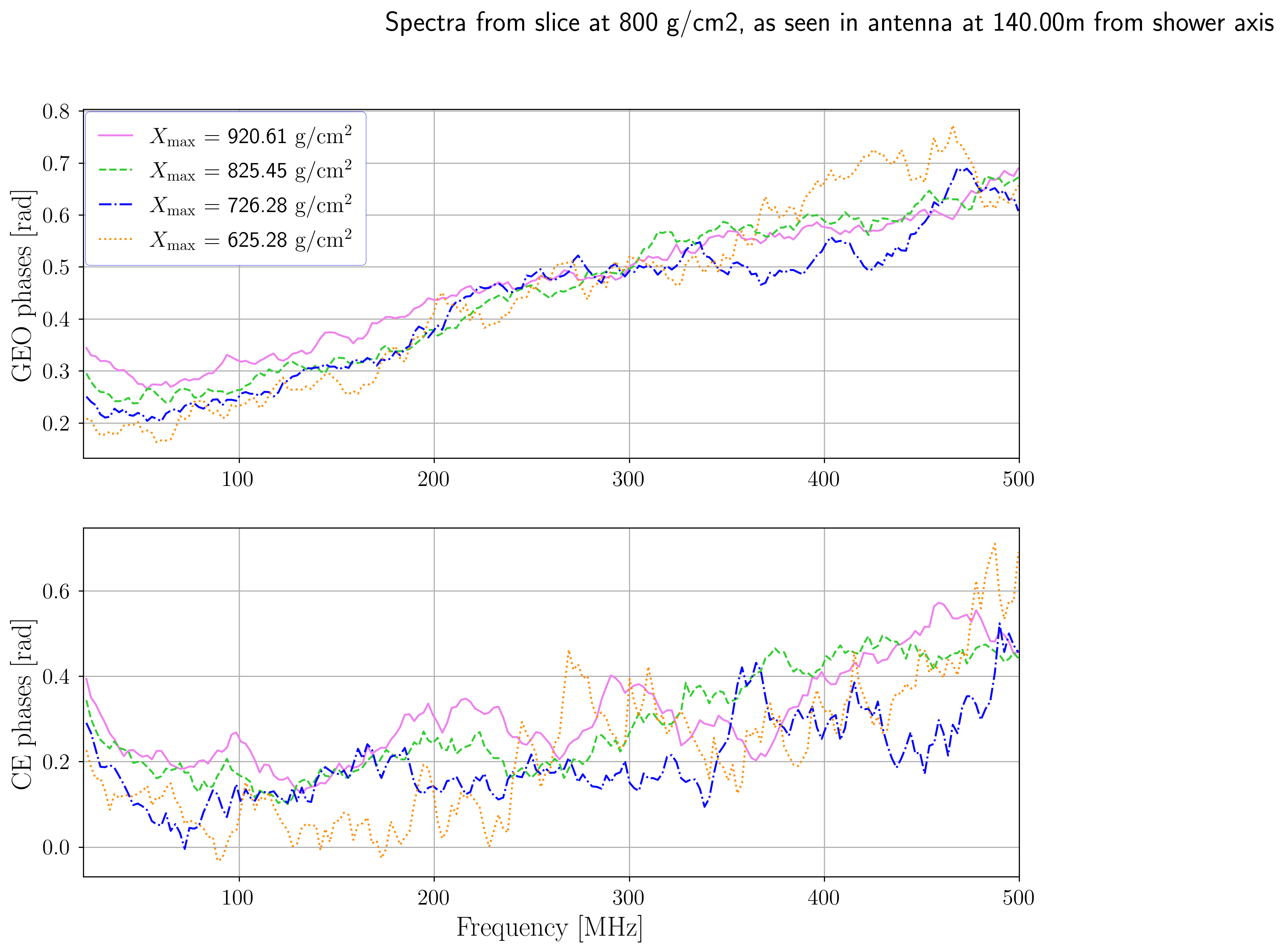}
        \label{fig:phases_multiple_showers}
    } \\
    \subfloat[Phase spectra in multiple antennas, from a single shower with \textXmax\ of 920 \textgcm . The labels in the legend are the viewing angles of the antennas, expressed as fraction of the Cherenkov angle of the slice.]{
        \includegraphics[width=0.6\linewidth, trim={0 0 3.5cm 2.7cm}, clip]{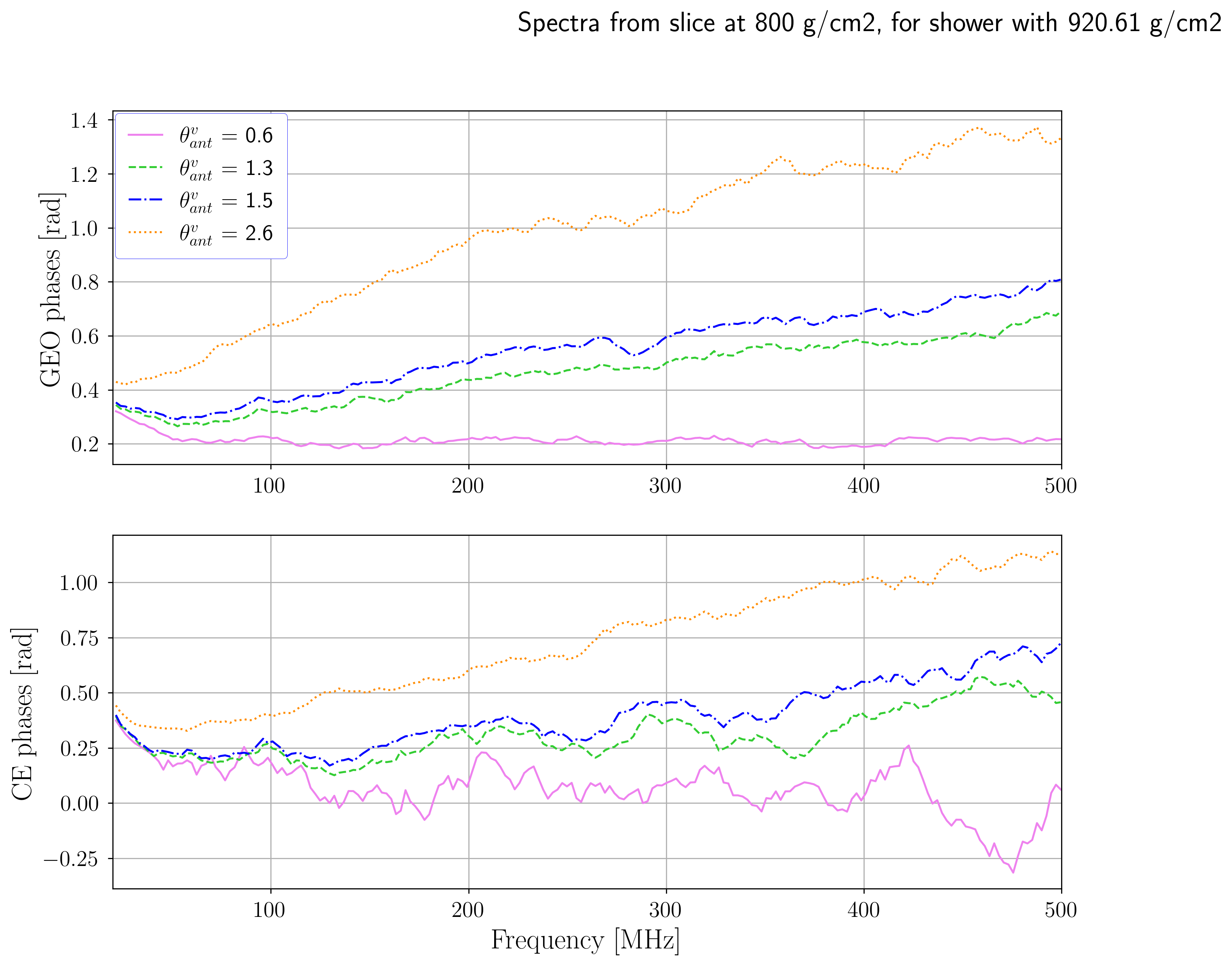}
        \label{fig:phases_multiple_ant}
    }
    \caption{
        Phases coming from a slice at 800 \textgcm , after unwrapping and correcting for the arrival time as per Equations \eqref{eq:arrival_time} and \eqref{eq:phase_shift}.
        There is little dependency of the spectra on the shower age, as seen in (a).
        The viewing angle of the antenna, whose effect is illustrated in (b) does have a important influence though.
        Therefore we always keep the viewing angle of the antennas fixed, instead of the position on the ground, when synthesising across air shower geometries.
    }
    \label{fig:phases_multiple}
\end{figure*}

We found that keeping the phases fixed to those of the origin shower resulted in quick degradation of performance when changing the zenith angle.
However, by simply accounting for the arrival time of the pulses based on the geometrical delays, the result of synthesis improved dramatically.
Even though the phase spectra do show deviations from perfect linearity (after unwrapping), as shown in Figure \ref{fig:phases_multiple}, it does appear that these deviations are relatively constant.

Our treatment of the phases is as follows.
Following the convention used in CoREAS, we take $t=0$ to be the time at which the air shower core hits the ground.
Considering a slice at grammage $X_{\slice}$, we can say that the time $t_{\text{emitted}}$ at which the signal from that slice was emitted is
\begin{align*}
    d \cdot \cos( \theta ) &= -c_i \log \left( \frac{X_{\slice} \cdot \cos(\theta) - a_i}{b_i} \right) \; , \\
    t_{\text{emitted}} &= - \frac{d}{c} \; .
\end{align*}
Here $\theta$ is the zenith angle, $c$ the speed of light in vacuum and the $a_i, b_i$ and $c_i$ are the parameters of the exponential profile\footnote{Our notation follows the convention used in \cite{Heck2023}, which can be confusing with our choice for the spectral parameters. Therefore we do want to point out these are different from the spectral parameters.} describing the atmosphere \cite{Heck2023}.
Because the shower front propagation is not slowed down by the index of refraction, we do not need to take it into account here.
Note that in calculating the distance we do assume a flat Earth geometry, which is why we can simply multiply by $\cos(\theta)$.
This is a very good approximation up to approximately 70\textdegree\ zenith angle.
The derivation of this formula is given in \ref{app:arrival_time}.

We then need to calculate the time $t_{\text{travel}}$ it takes for the signal to travel from the slice to the antenna.
With $d_{\slice}$ as defined in Figure \ref{fig:geometry}, we can simply say that
\begin{align*}
    n_{\text{eff}} &= 1 + \frac{n_{\text{sea}} - 1}{\rho_{\text{sea}}} \left[ T_{\ant} - T_{\slice} \right] \; , \\
    t_{\text{travel}} &= \frac{n_{\text{eff}} \cdot d_{\slice}}{c} \; .
\end{align*}
Here the subscript ``sea'' refers to values at sea level. 
The $n$ and $\rho$ are refractive index and density respectively.
The function $T$ gives the mass overburden at some height, in the above formula at the height of the antenna (observation level) and that of the slice.
In the above equations, the first line is essentially calculating the effective refractive index integrated along the line of sight from the slice to the antenna.
The given formula is an analytical approximation which is again valid for flat geometries, effectively up to about 70\textdegree\ zenith angle, where the slant depth can easily be related to the height above ground.\footnote{This integration as well as the previous distance calculation, are not implemented by hand. Rather we rely on the \texttt{radiotools} package (https://github.com/nu-radio/radiotools/) for this.}
A more detailed derivation is provided in \ref{app:arrival_time}.

Finally we have an expected arrival time in the antenna $t_{\text{arrival}}$, based on the air shower geometry.
\begin{equation} \label{eq:arrival_time}
    t_{\text{arrival}} =  \frac{n_{\text{eff}} \cdot d_{\slice} - d}{c} \; .
\end{equation}
We can adjust the pulse from a slice to remove this arrival time, i.e. moving the peak of the pulse to $t = 0$ by adding
\begin{align} \label{eq:phase_shift}
    2 \pi \cdot f \cdot t_{\text{arrival}}
\end{align}
to its phase spectrum.
In this equation $f$ refers to the values on the frequency axis after the Fourier transform.
The peak can then be moved back to the expected arrival time by subtracting the same amount.
If the origin and target shower have the same geometry, these two operations will cancel out.
However, if the target has a different geometry, the $t_{\text{arrival}}$ will be different and hence the peak of the pulse will be a different time.
This can have important consequences on the total pulse, as the difference in arrival time will be different for every slice.
Hence the signal from each slice will be shifted by a different amount, which will affect the coherency between them.

In Figures \ref{fig:phases_multiple_showers} we show the phase spectra from multiple showers in the same slice, as recorded in a single antenna.
The spectra have been unwrapped and the linear component from the arrival time, as given by Equation \eqref{eq:phase_shift}, has been subtracted for clarity.
We can appreciate the dispersion, i.e. the higher-order terms in frequency, as deviations from horizontal lines in these plots.
Furthermore, we can observe that the variations from showers with different \textXmax\ are $\lesssim 0.2$, which corresponds to shifts that are small compared to the wavelengths.
These are therefore too small to change the level of coherency between the slices.
In our usual interpretation of \textDXmax\ we conclude that the shower age has no effect on the phases.
This explains why we did not have to account for the phases when keeping the geometry fixed.

The viewing angle of the antenna on the other hand, has an important effect as we can see in Figure \ref{fig:phases_multiple_ant}.
Here we also removed the linear component due to the different arrival times in the antennas.
For this reason we always keep the viewing angle of the antennas fixed in template synthesis, even when changing geometry.

\subsection{Creating a template} \label{subsec:template}

The synthesis process can be divided in two steps.
The first one is to generate an input microscopic simulation, called the origin shower, and process it into a template.
We will discuss this procedure here.
In the next section we discuss how we synthesise the radio signals of new showers with user-defined longitudinal developments from these templates.

The origin shower is a microscopic simulation, sliced in atmospheric depth.
This simulation should contain all the antennas of interest, as the template will only be defined for those.
The reason for this is that the phases contain important information about the relation between slices and antennas, for which we rely on the origin shower.
If you would like to synthesise the emission at arbitrary positions on the ground, we suggest to use an origin shower with antennas arranged in a star-shape pattern, which can then be interpolated using for example the method described in \cite{corstanje_high-precision_2023} after synthesising the radio signal.

For every antenna, we loop over all slices.
Each trace $E_{\slice}$ is decomposed into the GEO and CE components.
For both components we calculate the amplitude frequency spectrum $A_{\origin}$, as well as the phase frequency spectrum $\phi_{\origin}$.
The amplitude frequency spectra are then normalised according to the slice and shower parameters, as we discussed in subsection \ref{subsec:scaling_relations}.
We then further normalise the amplitude spectra using the spectral functions, evaluated at the $\DXmax^{\slice}$ of the current slice and the viewing angle of the antenna.
The phase spectra are shifted with the arrival times calculated from the origin geometry, using Equations \eqref{eq:arrival_time} and \eqref{eq:phase_shift}.

The phase spectrum, together with the final normalised amplitude spectrum per slice, constitute the template for one antenna.
We have these for each antenna in the origin shower.
Combining all antennas together yields what we will refer to as the template in the following.
The template only needs to be constructed once for a given origin shower.
It can be stored and reused later.

\subsection{Applying template synthesis to retrieve the radio signal} \label{subsec:synthesis}

Equipped with a template, we can now synthesise the emission from an air shower with a different longitudinal distribution and, if desired, modified arrival direction (to within a few degrees).
We call this the target shower.
In each antenna, we again loop over all the slices in the template.
We now evaluate the spectral functions with the $\DXmax^{\slice}$ of the slice, which is now calculated with the target \textXmax .
This is multiplied with the template amplitude spectrum.
We then also undo the geometry normalisation from subsection \ref{subsec:scaling_relations}, using the values from the target shower.

The template phase spectrum is also adjusted using the geometry of the target shower, as follows.
We first calculate the arrival times from the signals of each slice in every antenna using Equation \eqref{eq:arrival_time}.
Then we use Equation \eqref{eq:phase_shift} with the opposite of these arrival times (i.e. multiplied by -1) to move the pulses to the expected peak times.
If the target has the same geometry as the origin shower, this will effectively cancel out the change made in the previous step.
But if this is not the case, the pulses will be shifted in time with respect to the origin.
This shift will be different in every slice as the changes in arrival time depend in a non-linear way on the zenith angle.
Hence the coherency between the different slices will be affected.

The resulting amplitude spectrum is recombined with the phase spectrum to produce the electric fields for both the GEO and CE components.
After summing up the contributions from all slices, we are left with two electric field traces in each antenna.
These can be recombined into a three-dimensional electric field using the polarisation patterns in the shower plane, assuming there is no emission in the component along the shower axis.

A practical overview of generating and mapping a template shower using the \texttt{SMIET} software we created is given in Section \ref{sec:workflow}.

\section{Building the \texttt{SMIET} package} \label{sec:workflow}

Up until this point, we only talked about theory behind template synthesis.
In this section we put it into practice and construct the software package that implements the templates synthesis algorithm.
We are delighted to open-source this Python package under the GPLv3.0 license, such that the entire astroparticle community can benefit from it.
As of May 2025 it lives at \cite{smiet_cr_synthesis}.
We refer to the documentation for instructions on how to install and use it.

The key component of the \texttt{SMIET} package are the spectral functions.
They are provided as a set of $p$-parameters, stored in a file which is part of the repository.
Hence a user is not required to generate those themselves.
In subsection \ref{subsec:extracting} we explain how we used a simulation library of almost 800 showers to extract them, using the method described above.
All these showers were simulated with the same atmosphere, refractive index profile, observation level and azimuth angle.
Then in subsection \ref{subsec:user_perspective} we describe the typical workflow of a user of our software.

\subsection{Obtaining the spectral functions} \label{subsec:extracting}

To extract the spectral functions we used a set of microscopic simulations, using CORSIKA v7.6400 with QGSJET-II and FLUKA as interaction models.
The choice of hadronic interaction model is not relevant for the following however, as the radio emission is purely driven by the electromagnetic cascade.
The thinning level is $10^{-7}$, with optimised weight limitation.
For the hadrons and muons we use a low-energy cutoff of $0.3 \GeV$, while for electrons and photons this value is $0.4 \GeV$.
We use the magnetic field vector from the LOFAR site, which has a strength of 0.492 Gauss and an inclination of 67.8 degrees.
The refractive index is modelled using the Gladstone-Dale model, with the value at sea level being set to $n = 1.000292$.
The atmospheric model we use is the US standard atmosphere parametrised by Keilhauer, which is model 17 in CORSIKA \cite{Heck2023}.

We made several simulation sets, each with a different zenith angle. 
The azimuth angle is always fixed to 0 degrees, which in the CORSIKA coordinate system means the particle comes in from the South.
This should not limit the generality of our results, as the azimuthal angle only influences the geomagnetic angle and hence the relative scaling between the geomagnetic and charge-excess components.
For each simulation set, we took half of the showers to be initiated by proton and the other half to be initiated by an iron nucleus.
The energies were randomly sampled from a log-uniform distribution between $10^{17} \eV$ and $10^{19} \eV$.
In \cite{desmet_proof_2024} we already concluded that we can safely correct for energy effects over two orders of magnitude.
In Table \ref{tab:simulations} we give an overview of the different simulation sets.

\begin{table*}
    \centering
    \captionsetup{width=.8\textwidth}
    \caption{
        Overview of the simulation sets used to create the spectral functions for the \texttt{SMIET} software package.
        We use the same set to also benchmark the performance of the software in Section \ref{sec:benchmarking}.
    }
    \begin{tabular}{l c c c c c}
        \hline
        \textbf{Zenith angle [deg]} & 20 & 30 & 40 & 50 \\
        \hline
        \hline
        \textbf{Number of simulations (p + Fe)} & 200 & 199 & 199 & 200 \\
         \hline
         \textbf{Atmospheric depth at ground [\textgcm]} & $1100.15$ & $1193.74$ & $1349.54$ & $1608.32$ \\
         \hline
         \textbf{Number of slices (using 5 \textgcm )} & 221 & 239 & 270 & 322 \\
         \hline
         \textbf{Number of simulated antennas} & 32 & 10 & 49 & 49 \\
         \hline
    \end{tabular}
    \label{tab:simulations}
\end{table*}

For each zenith angle, we choose a set of antenna positions on the $\vvB$ axis.
Their distances to the shower axis were chosen to span from half up to three times the typical Cherenkov radius of the geometry in question, with the distance range near the Cherenkov cone itself being sampled more densely.
With ``typical'' we mean the Cherenkov radius for an average \textXmax\ at $10^{18} \eV$, given the zenith angle.

In CoREAS, we configure the observers with a sampling rate of $0.2 \ns$ and a time window of $400 \ns$.
When slicing a simulation in CoREAS, the number of observers is essentially multiplied by the number of slices.
For example, showers from the 50\textdegree\ zenith angle set from Table \ref{tab:simulations} had the equivalent of 9800 observers.
Combining this with the very fine sampling in time, yields runtimes on the order of weeks.
In order to keep them reasonable we simulated the showers multiple times, each time with a different subset of antenna positions.

In the following we will want the antennas at fixed viewing angles, for each slice.
To achieve this with our simulations, we used the Fourier interpolation method described in \cite{corstanje_high-precision_2023}.
For every slice we interpolated the electric field timeseries on the $\vvB$ axis to the viewing angle that we require.

In order to obtain the spectral functions, we apply the same procedure to all simulations.
We present the pseudo-code of this procedure in Algorithm \ref{algo:spectral_functions}.
First we define a list of normalised viewing angles for which we want to find the spectral functions.
These were chosen such that close to the Cherenkov angle (which has a viewing angle of 1, as this is given as a fraction of the Cherenkov angle) we have many samples, while still having a wide range of interesting angles.

For each viewing angle, we then loop over every slice that we simulated.
We start by interpolating the signal to the viewing angle we are considering.
Then we decouple the GEO and CE emission components.
Because all antennas are on the $\vvB$ axis, this is readily done by selecting one polarisation of the electric field.
After this we calculate the amplitude frequency spectrum, by applying the real-valued Fourier transform and taking the absolute value.
We then fit the geomagnetic and charge-excess components using Equations \eqref{eq:a_geo} and \eqref{eq:a_ce}, respectively.
These include the normalisation discussed in subsection \ref{subsec:scaling_relations}.
The values of the fitted spectral parameters are then stored, together with the $\DXmax^{\slice}$ of the slice and the chosen viewing angle.

\begin{algorithm}
\caption{Extracting the spectral functions for the \texttt{SMIET} software} \label{algo:spectral_functions} 
\begin{algorithmic}
    \Require \textit{dva} = desired viewing angles
    \ForAll {shower}
    \ForAll {slices}
    \ForAll {angle $in$ \textit{dva}}
        \State ant = \verb|interpolate_to_viewing|(angle)
        \For { comp = GEO, CE }
            \State trace =  \verb|get_trace|(ant, slice, comp) 
            \State norm = \verb|remove_scalings|(trace, comp)
            \State amp = \verb|abs| ($RFFT$(norm))
            \State a, b, c = \verb|fit_amp_spectrum|(amp, comp)
            \State database $\gets$ angle, a, b, c, $\DXmax^{\slice}$
        \EndFor
    \EndFor
    \EndFor
    \EndFor
\algstore{step1}
\end{algorithmic}
\end{algorithm}

We then proceed to the construction of the spectral functions, which also happens for each viewing angle separately, as described in subsection \ref{subsec:spectral_functions}.
For each angle, we bin the fitted values of the spectral parameters in $\DXmax^{\slice}$.
To the mean and standard deviation in each bin, we then fit a quadratic function, as demonstrated in Figure \ref{fig:spectral_fits}.
This finally yields us the five different spectral functions, two for the CE component and three for the GEO one, per viewing angle that we defined.

\begin{algorithm}
\begin{algorithmic}
    \algrestore{step1}
    \ForAll {angle $in$ \textit{dva}}
        \State $a, b, c, \DXmax^{\slice} \gets$ \verb|find_vars|(database, angle)
        \ForAll {spectral $\in ( a_{geo}, b_{geo}, c_{geo}, a_{ce}, b_{ce} )$}
            \State $\text{p}^i$ =  \verb|parabola_fit|($\DXmax^{\slice}$, spectral)
            \State \verb|spectral_functions[angle]| $\gets$ $\text{p}^i$
        \EndFor
    \EndFor
\end{algorithmic}
\end{algorithm}

These spectral functions are each encoded using 3 $p$-parameters.
It is those values that we store to disk and include in the \texttt{SMIET} software package.
The $p$-parameters depend on the frequency range over which the spectra were fitted.
They need to extracted separately for each frequency range of interest.
In the software package we provide the spectral functions for the [30, 80] \textMHz\ and [30, 500] \textMHz\ frequency bands.

\subsection{A user's perspective} \label{subsec:user_perspective}

\begin{figure}
    \centering
    \begin{tikzpicture}[node distance = 2cm, auto]
        \node [start] (set) {Choose \textXmax\ range of interest and EAS geometry};
        \node [cloud, below of=set] (produce) {Produce input CORSIKA files};
        
        \node [block, below of=produce] (init) {Simulate origin shower(s)};
        
        \node [cloud, below of=init] (spectra) {Process origin into a template};
        \node [decision, right of=spectra, node distance=4.2cm] (save) {Save template to disk};
        
        \node [start, below of=spectra, text width=11em] (stop) {Choose the longitudinal profile of target shower};
        \node [cloud, below of=stop] (fit) {Map template to target shower};
        
        \path [line] (set) -- (produce);
        \path [line] (produce) -- (init);
        \path [line] (init) -- (spectra);
        \path [line] (spectra) -- (stop);
        \path [line] (stop) -- (fit);
        \path [line, dashed] (spectra) -- (save);
    \end{tikzpicture}  
    \caption{
        Typical user workflow for the template synthesis software.
        The steps highlighted in red ellipses (as well as the save operation in the diamond shape) are performed using functions in the package.
        After choosing a range of interest, an appropriate origin shower needs to be simulated with CoREAS.
        If the range in \textXmax\ is larger than 200 \textgcm , multiple origin showers should be simulated for the best accuracy.
        After reading in the sliced origin simulations and processing them into a template, they can optionally be saved to disk.
        Since this processing step is computationally more expensive than the mapping, saving the template is recommended.
        Once a template has been loaded (either from computation or from disk), the longitudinal profile of the target shower has to be chosen.
        This feeds into the algorithm to synthesise the radio emission.
    }
    \label{fig:workflow_diagram}
\end{figure}
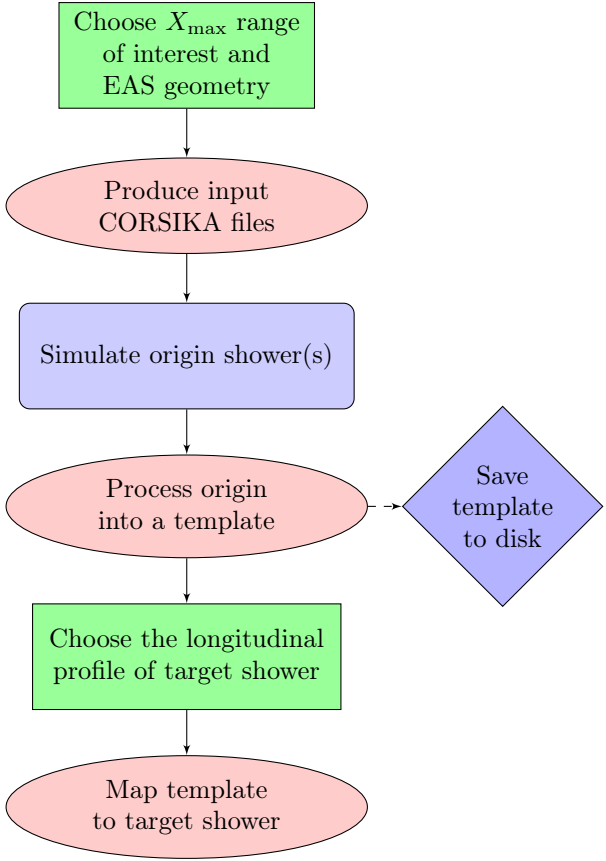

The \texttt{SMIET} software has been designed to get new users up and running with template synthesis as fast as possible.
Several classes are provided to facilitate the interaction with the files and to run the algorithm \cite{smiet_cr_synthesis}.
As of the publication of this article, the spectral functions for two frequency ranges are provided as part of the package: [30, 80] \textMHz\ and [30, 500] \textMHz .
In order to use template synthesis with other frequency ranges, you can synthesise traces over the broad frequency range and filter down after.

In Figure \ref{fig:workflow_diagram} we sketch the typical, basic user workflow.
The steps highlighted in red can be achieved with a single function call in the software.
Generating the template takes quite a bit more time than mapping it.
This is why we recommend saving the template to disk once it is created, which can also easily be done using a single function call in the software.
Once the template has been generated, the mapping part only takes a couple of seconds.

Next to the code necessary to run the synthesis, the package also offers functionality to generate the CORSIKA input files to simulate origin showers.
These take care of adding enough slices and also automatically configure the settings to run with MPI.
The antennas can be either placed at the same position for every slice, which is the recommended setting for generating showers, or they can be configured to be placed under the same set of viewing angles for each slice.
The latter option can be used for experimentation and to regenerate the spectral parameters from scratch.

Because template synthesis relies on the decoupling of the GEO and CE components, we suggest to only work with antennas on the $\vvB$ axis.
There, the decoupling is the easiest and most reliable.
After performing synthesis on these antennas, a complete star shape can be reconstructed from the known polarisation patterns in the shower plane.

In the future, we plan to provide an interface in the NuRadio framework \cite{glaser_nu-radionuradiomc_2025}.
This should make it straightforward to integrate template synthesis into analyses for which CoREAS is currently used.
Together with this interface we also plan to create a library of origin showers that can be used by everyone.

\section{Benchmarking the performance} \label{sec:benchmarking}

With the \texttt{SMIET} software package ready to be used, we now want to verify its accuracy.
For this we benchmark template synthesis in subsection \ref{subsec:performance} by synthesising microscopic simulations onto each other.
This allows us to explicitly compare the results of template synthesis to full CoREAS simulations. 
We will use the same simulation set used to extract the spectral functions, which is described in subsection \ref{subsec:extracting}. 
Since the benchmarks show that template synthesis has some biases, we introduce in subsection \ref{subsec:interpolated_synthesis} an interpolation approach which deals with them in an elegant way. 
From these benchmarks we then conclude some guidelines for the practical application of our method in subsection \ref{subsec:guidelines}.

\subsection{Performance metrics over the extraction set} \label{subsec:performance}

In order to quantify the quality of a synthesised signal, we define several metrics.
Here we compare the electric field trace $s_{\real}$ produced by CoREAS and decomposed into the GEO and CE components using the polarisation patterns in the showerplane, to the signal $s_{\synth}$ synthesised using our method with the exact longitudinal profile as given by CORSIKA.

The first one is the ratio of the maxima of the two time traces,
\begin{equation} \label{eq:peak_ratio}
    S_{\text{peak}} (s_{\synth}, s_{\real}) = \frac{\max (s_{\synth}(t))}{\max (s_{\real}(t))} \; .
\end{equation}
We will refer to this quantity as the peak ratio in the following.
Because taking the maximum of a very noisy trace can result in a random value at the end of the trace, we limit the time range over which the maximum can be selected.
In order to do this, we first look for the peak amplitude in the time trace after filtering it down to [20, 100] \textMHz . 
In this band the signal peak is typically much more defined.
We define a window of $\pm 5 \ns$ around the time of this peak.
To then calculate the peak ratio, we look for the maximum in the full-bandwidth trace within this determined window.
This is done for the synthesised and CoREAS traces separately.

The peak ratio relates closely to another important quantity called the energy ratio, namely the ratio of the energy fluence contained in the signals,
\begin{equation} \label{eq:energy_ratio}
    S_{\text{fluence}} (s_{\synth}, s_{\real}) = \frac{\mathcal{F} (s_{\synth}(t))}{\mathcal{F} (s_{\real}(t))} \; ,
\end{equation}
where the fluence is calculated as
\begin{equation}
	\mathcal{F}(s) \propto \Delta t \sum_{m} s(m \cdot \Delta t)^2 \; .
\end{equation}
Here $\Delta t$ is the sample time used in the simulation and $m$ runs over the samples.
Also here we use the peak find algorithm mentioned above, to avoid integrating over the noisy parts of the trace.
We calculate the fluence in a window of $\pm 25 \ns$ around the peak time.

Another interesting quantity is the time offset between the maxima.
This can be determined using the Hilbert envelope $\mathcal{H}$, which is found as
\begin{align*}
    \mathcal{H}(s(t)) = | FFT^{-1} (FFT(s) 2 U) | \; ,
\end{align*}
where $FFT$ indicates the Fourier transform and $U$ is the unit step function.
To obtain the time offset, we can look at the difference in timing between the peaks of the Hilbert envelopes of the synthesised and CoREAS signal,
\begin{align} \label{eq:hilbert_max}
    S_{\text{Hilbert}} (s_{\synth}, s_{\real}) &= \nonumber \\
    t|_{\max (\mathcal{H} (s_{\synth}))} &- t|_{\max (\mathcal{H} (s_{\real}))} \; .
\end{align}
The reason to use the Hilbert envelope instead of the raw signal, is that the former does not depend on where in the time bin the actual pulse is.

\begin{figure*}
    \centering
    \includegraphics[width=\linewidth]{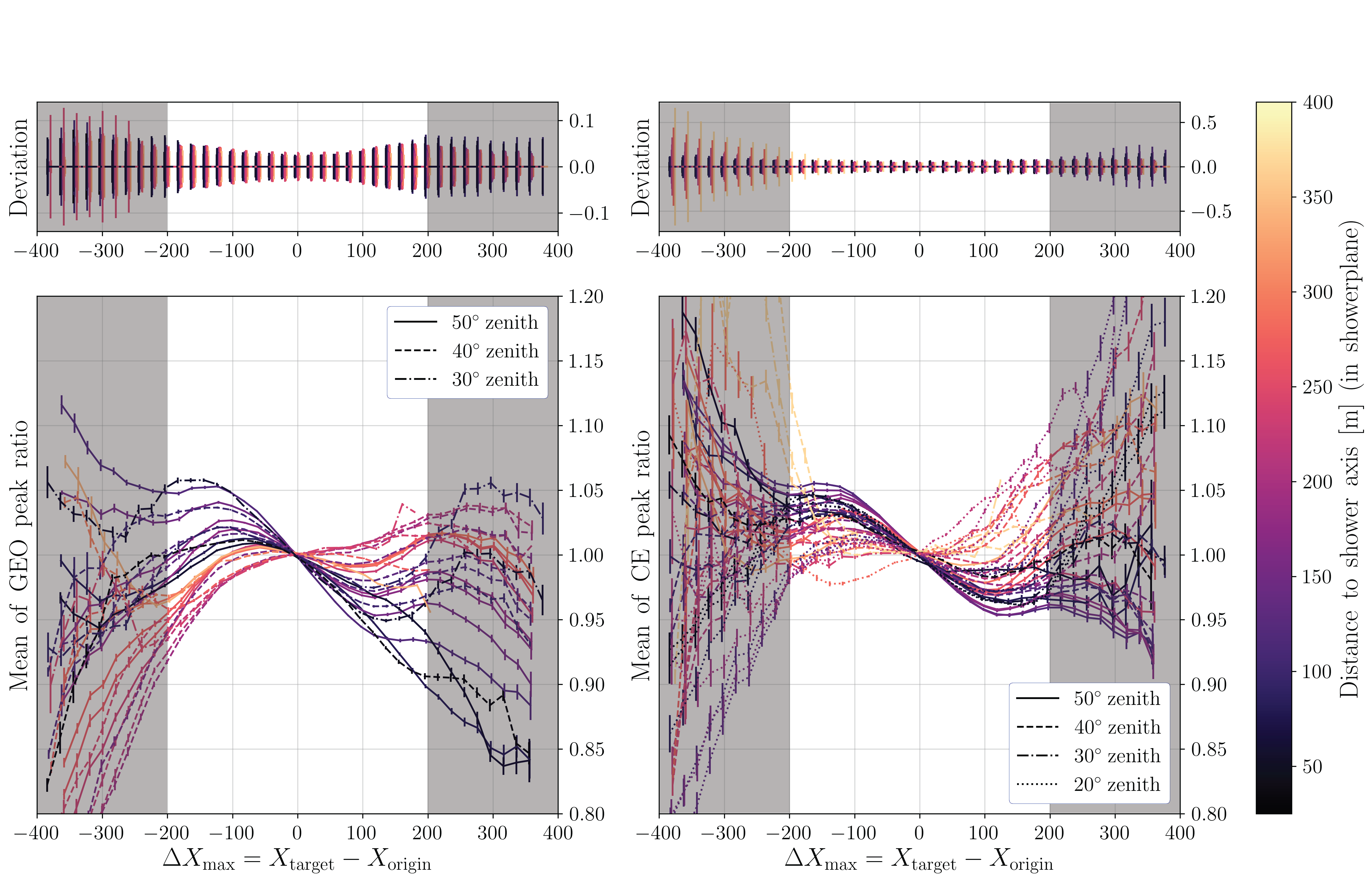}
    \caption{
        The peak ratio metric $S_{peak}$ from Equation \eqref{eq:peak_ratio}, calculated over the simulation sets detailed in Table \ref{tab:simulations}. 
        The sets are differentiated using different linestyles.
        The colour of the lines reflects the distance to the shower axis in the shower plane.
        In these plots, the values of the metric are binned in \textDXmax .
        The plots at the top show the standard deviation in every bin, where the entries are slightly shifted within each bin for better clarity.
        Underneath those we show the evolution on the mean value in each bin.
        Here the error bars are the standard deviations divided by the square root of the number of samples in the bin to reflect the uncertainty on the mean estimation.
        These results were obtained by synthesising over the [30, 500] \textMHz\ band.
        To create these figures, we only included antennas which had at least 5\% of the peak amplitude found in the set, to exclude noisy antennas with little signal.
        In the plot of the geomagnetic component we removed the 20 \textdegree\ shower set because its geomagnetic signal is too small and the resulting fluctuations are mainly due to numeric noise.
        For the other three sets the deviations are about 2\% around $\DXmax = 0 \gcm$ and grow up to 6\% at $\DXmax = \pm 200 \gcm$.
        These values are similar for all antennas and zenith angles.
        This is not the case for charge-excess, where the antennas closest to the shower axis have the biggest deviations.
        Since the charge-excess contribution drops to zero near the shower axis, this is to be expected.
        The antennas where the signal is stronger show a similar behaviour as the geomagnetic case.
        The standard deviations are the smallest around $\DXmax = 0 \gcm$, about 4\%.
        In this case they grow up to 9\% at $\DXmax = \pm 200 \gcm$.
        In general the results show the biggest standard deviations for the weakest parts of the signal, where we can expect there be more scatter inherently.  
        Overall we can conclude that the safe range for application of template synthesis is $|\DXmax| \leq 200 \gcm$, which we indicate by the grey bands in the plots.
    }
    \label{fig:benchmarks}
\end{figure*}

The results of the peak ratio benchmarks is shown in Figure \ref{fig:benchmarks}.
Since the conclusions from the other benchmarks align with the following, we refer readers to \ref{app:more_plots} to see the plots for the energy ratio and comparison of the Hilbert timings.
We used the simulations sets given in Table \ref{tab:simulations}.
The benchmarking was performed per simulation set, i.e. per zenith angle.
Within each set, we took every shower as the origin to synthesise all other ones.
In each antenna we then calculated the peak and energy ratio scores, and recorded those with the \textDXmax\ of the two.
To understand how the scores change with this parameter, we bin them in \textDXmax .

Since the simulation set contains showers with strongly varying footprint sizes, we only select antennas which are contained in the footprint by applying a cut on the peak amplitude.
For each shower we find the peak amplitude of both components in each antenna using the CoREAS traces.
If a trace has less than 5\% of the highest recorded peak of that component in the simulation, it is not considered for the benchmarking.\footnote{This does not mean that 5\% of the antennas is removed from the set. Depending on the peak values in each antenna, this criteria could remove none of them or almost all.}
After this cut we can then calculate the mean and the standard deviation in each bin.
This is what we show in Figure \ref{fig:benchmarks}.

In the bottom plots, which show the mean value per bin, each coloured line refers to one antenna.
The uncertainty on the mean value is the standard deviation divided by the square root of points in the bin.
The lines are colour-coded by the distance to shower axis, in the shower plane.
Overall we observe biases typically below 5\%, with a few antenna locations going up to 10\% over the range $\DXmax \in [-200, 200] \gcm$.
These have a particular shape and become smaller as the \textDXmax\ becomes smaller.
The parameter which correlates most strongly with this bias is the distance of the antenna to shower axis in the shower plane. 
It is interesting to note that the bias seems to be very symmetrical around $\DXmax = 0$.
We suspect the shape of the bias reveals something about its origin, but so far we have not been able to track it down.
Therefore we have to postpone this to future work.
We could opt to fit this bias and correct for it, akin to the approach we took in \cite{desmet_proof_2024}, but we prefer to not complicate the method any further.
Instead, we will present a different approach which can more naturally correct for this bias in the next section.

Still, we can appreciate that for the antennas which are close to the typical Cherenkov ring (150 - $300 \m$), the synthesis performs better.
This probably hints at the fact that the biggest deviations are from traces which simply do not contain a lot of signal, namely the antennas very far away and very close to the shower axis (the latter in particular for the charge excess emission).
We presume that the shape of the amplitude frequency spectrum will also play a big role.
Close to the Cherenkov ring, the spectrum becomes very flat which is fitted more easily.

We can also see that the different lines overlap, irrespective of the zenith angle of the set.
This is a good indication that the method is indeed universal with respect to the geometry.
Even the biases seem to follow the same trend for all shower sets.

On top of the mean plots we show the standard deviation in each bin, not divided by the number of entries.
To limit the overlap, the different lines have been slightly shifted with respect to each other.
They are again coloured by the distance to the shower axis.
For the GEO component we can see that the standard deviation is minimal at $\DXmax = 0 \gcm$ and increases as the absolute value of \textDXmax\ increases.
This is the same behaviour as we observed in \cite{desmet_proof_2024} and is to be expected, as a larger \textDXmax\ is an indication that the origin and target showers are more different.
We do have to note that we had to remove the 20\textdegree\ zenith set from this benchmark, because it happened to have a very small geomagnetic angle of about 2.3\textdegree , which introduced large scatter in the GEO scores. 
We can also observe that the standard deviations are not dependent on the distance to the shower axis.

This is not the case for the CE component.
We can clearly observe that the standard deviations of the CE component become very large for antennas closer than $50 \m$ to the shower axis.
However, in this range, we do not expect a large contribution from this component \cite{Glaser2016}.
So we can again presume this is mostly due to the calculation of ratios of small numbers.
Close to the typical Cherenkov ring the standard deviations are minimal and they grow slightly again as we move to farther distances.
This seems to follow the trend of expected signal strength of the CE component.
Furthermore, since the CE is usually much weaker than the GEO component (as is clearly visible in Figure \ref{fig:phases_multiple}), we can presume the worse scores overall are due to a lack of clear signal, which would not be used for reconstructions anyway. 

\subsection{Interpolated synthesis to remove the bias} \label{subsec:interpolated_synthesis}

Since the biases in the benchmarks look symmetrical in \textDXmax , there is a straightforward way to compensate for them.
If we used two origin showers, one with an \textXmax\ bigger and one with an \textXmax\ smaller than the target, both synthesised traces should have opposite biases.
Hence, we could take the weighted mean of these two synthesised traces to get an unbiased result.
Since this amounts to performing a linear interpolation per time sample, we call this ``interpolated synthesis''. 

\begin{figure}
    \centering
    \includegraphics[width=1.0\linewidth, trim={0 0 0 2.5cm}, clip]{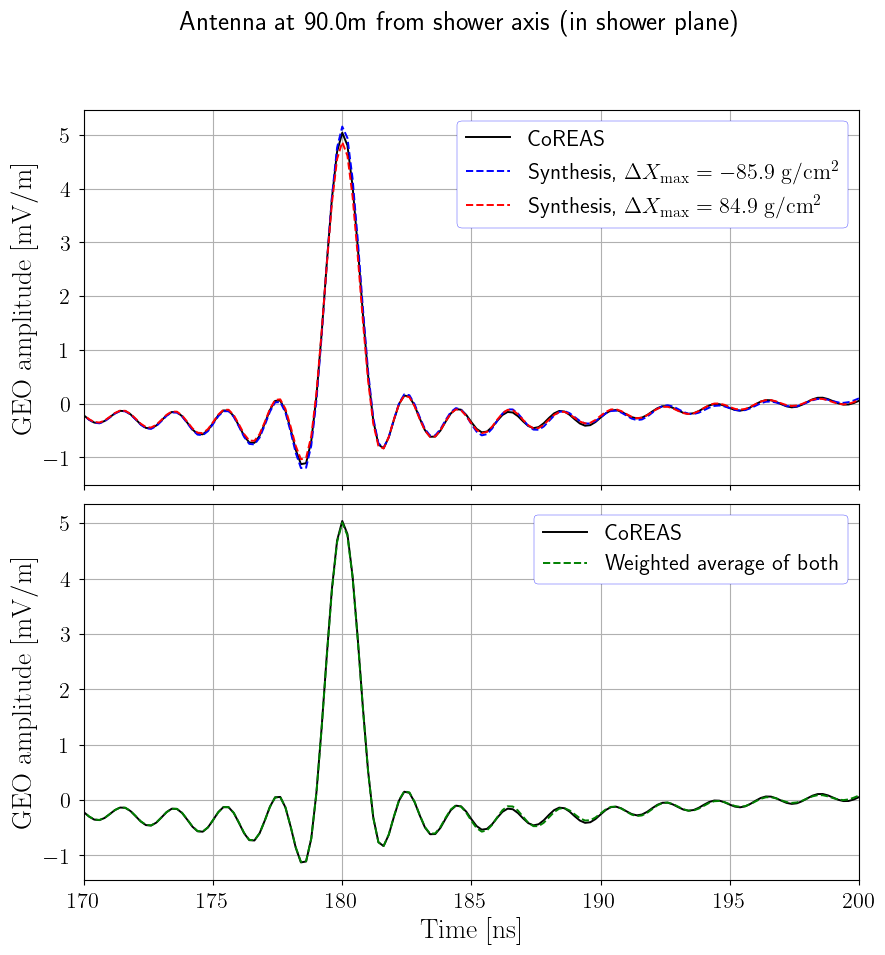}
    \caption{
        An example of applying a linear interpolation to the results of two syntheses, one with a positive \textDXmax\ and with a negative one.
        Here we show the electric field amplitudes of the geomagnetic component in an antenna at $90 \m$ from the shower axis (in the shower plane).
        For comparison we show the results of the individual syntheses in the top plot and their average in the bottom plot.
        We can see the biases in the top plot, where the blue line (negative \textDXmax) slightly overshoots the CoREAS result and the red line (positive \textDXmax) slightly undershoots.
        Taking the weighted average gives a result which traces the CoREAS line much better.
    }
    \label{fig:interpolated_synthesis}
\end{figure}

In Figure \ref{fig:interpolated_synthesis} we show an example of applying this procedure.
We can see that taking the weighted average indeed corrects for the biases and results in a synthesised trace which is almost perfectly in line with the CoREAS simulation.
In our example the two origin showers have similar, but opposite, \textDXmax\ with respect to the target.
This is only for better clarity, as the synthesised traces will have similar biases.
It is not necessary in general, as one can weight the contributions from both depending on the ratio of the two \textDXmax .

Apart from correcting for the biases, this approach also allows to construct an object which can synthesise over the whole \textXmax\ range of interest.
If, for example, a user wants to synthesise showers with an \textXmax\ between two known values, they could generate origin showers with \textXmax\ at intervals of 100 \textgcm .
This ensures that the target shower is always within 100 \textgcm\ of two origin showers, where the linear relation in the bias is the strongest.
For any target they would then pick the origin shower with the \textXmax\ above and below the target one, and use those two to perform the interpolated synthesis.

We have not investigated this avenue in much detail, so improvements to this approach will surely be possible.
These would probably take special care to adjust the interpolation for the non-linearities in the biases.
The shapes of the bias curves also seem to change in a consistent way with distance to the shower axis, which could be included as well.

\subsection{Using template synthesis in practice} \label{subsec:guidelines}

From the previous benchmarks, we can conclude a couple of things.
First of all, we see that as long as we stay with $100 \gcm$ of the origin shower, the peak amplitude of the synthesised geomagnetic component signal is within 4\% of the CoREAS signal over the frequency band [30, 500] \textMHz .
Going to a difference of $200 \gcm$, this difference increases to about 6\%. 
For the synthesised charge-excess component on the other hand, we observe a 6\% variation in the $\pm 100 \gcm$ range and a 9\% variation in the $\pm 200 \gcm$ range.
The inherent shower-to-shower fluctuations are bound to lead to scatter on the 4\% level \cite{desmet_proof_2024}.
They act as a lower bound on the accuracy our method will achieve, in the sense that we do not expect our benchmarks to have variations lower than this value.
As such we conclude that the template synthesis algorithm itself does not introduce much additional scatter.

In the end we can conclude that for practical applications of template synthesis, one should use origin showers whose \textXmax\ value is within $200 \gcm$ of the target \textXmax .
If a higher precision is required, the range should be limited to $100 \gcm$.
We also suggest users to adopt the interpolated synthesis approach, as it can be implemented without much effort and improves the results dramatically.

\section{Checking the universality of our approach} \label{sec:universality}

Until now, all simulations used to build the spectral functions and test the method were set to the magnetic field at the LOFAR site, an observation level at sea level, and the US standard atmosphere.
In order to see if our new approach is truly general, we need to verify that our spectral functions do not depend on these specific settings.
Therefore, in this section we will vary these and check if the performance is similar to the results in the previous section.
At the same time, these tests will also highlight several use cases for template synthesis.

For each analysis, we simulated 20 (sliced) showers, each with 26 antennas on the $\vvB$ axis.
In this section we used CORSIKA v7.7550, which includes a bugfix allowing us to use the MPI-enabled version.
This means we do not longer have to rerun simulations with subsets of antennas.
All other settings, such as interaction models, weight limitation, etc. are identical to subsection \ref{subsec:extracting}.
We perform the same benchmarks as in the previous section on this smaller set of showers, also applying the 5\% peak amplitude cut.
Of course, the statistics will be much lower, but we should still be able to observe the same trends in the data.
We still use the spectral functions extracted in Section \ref{sec:workflow}, but use the newly generated simulations as input showers.

\subsection{Applying template synthesis to the AERA site} \label{subsec:aera_showers}

One of the primary motivations to develop a more general template synthesis model is to be able to apply the approach in the context of many different experiments around the world. 
Therefore, in a first study, we looked at using simulations from a different experiment site.
We chose to make a shower set using the parameters of the AERA site \cite{krause_aera_2014}.
This includes setting the observation level to 1564 \textm\ above sea level, which will effectively reduce the number of slices we will process, as well as changing the magnetic field to a value of 19.69 $\mu \text{T}$ in the northern direction and 14.14 $\mu \text{T}$ in the upwards direction.
We did not change the atmospheric model and fixed the arrival direction to coming from the South with a 40\textdegree\ zenith angle.
The antennas are also still placed on the $\vvB$ axis.

\begin{figure*}
    \centering
    \subfloat[Geomagnetic component]{
        \includegraphics[height=8.5cm, trim={17.1cm 0 3.5cm 2.5cm}, clip]{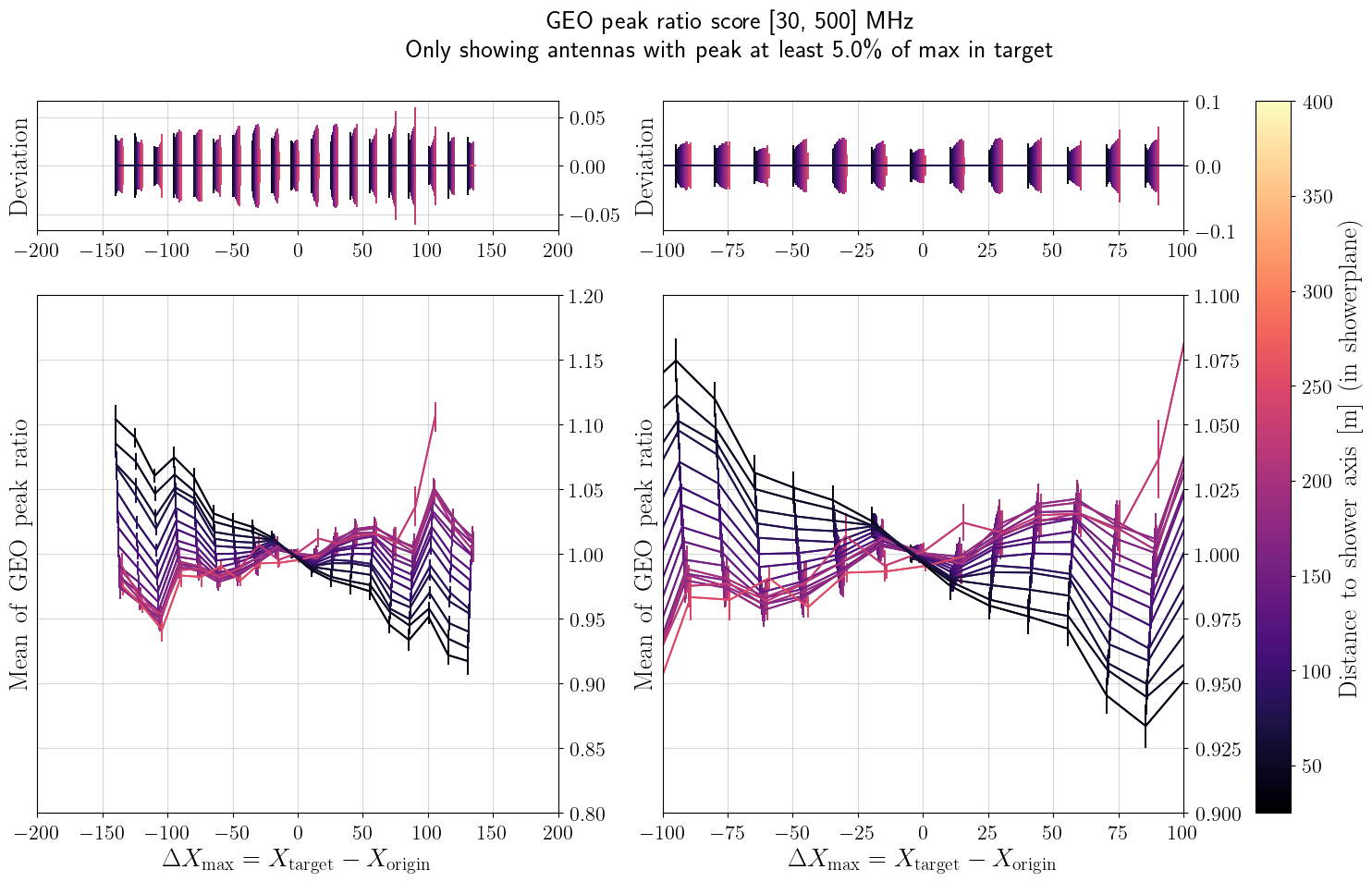}
    }~
    \subfloat[Charge-excess component]{
        \includegraphics[height=8.5cm, trim={17.1cm 0 0 2.5cm}, clip]{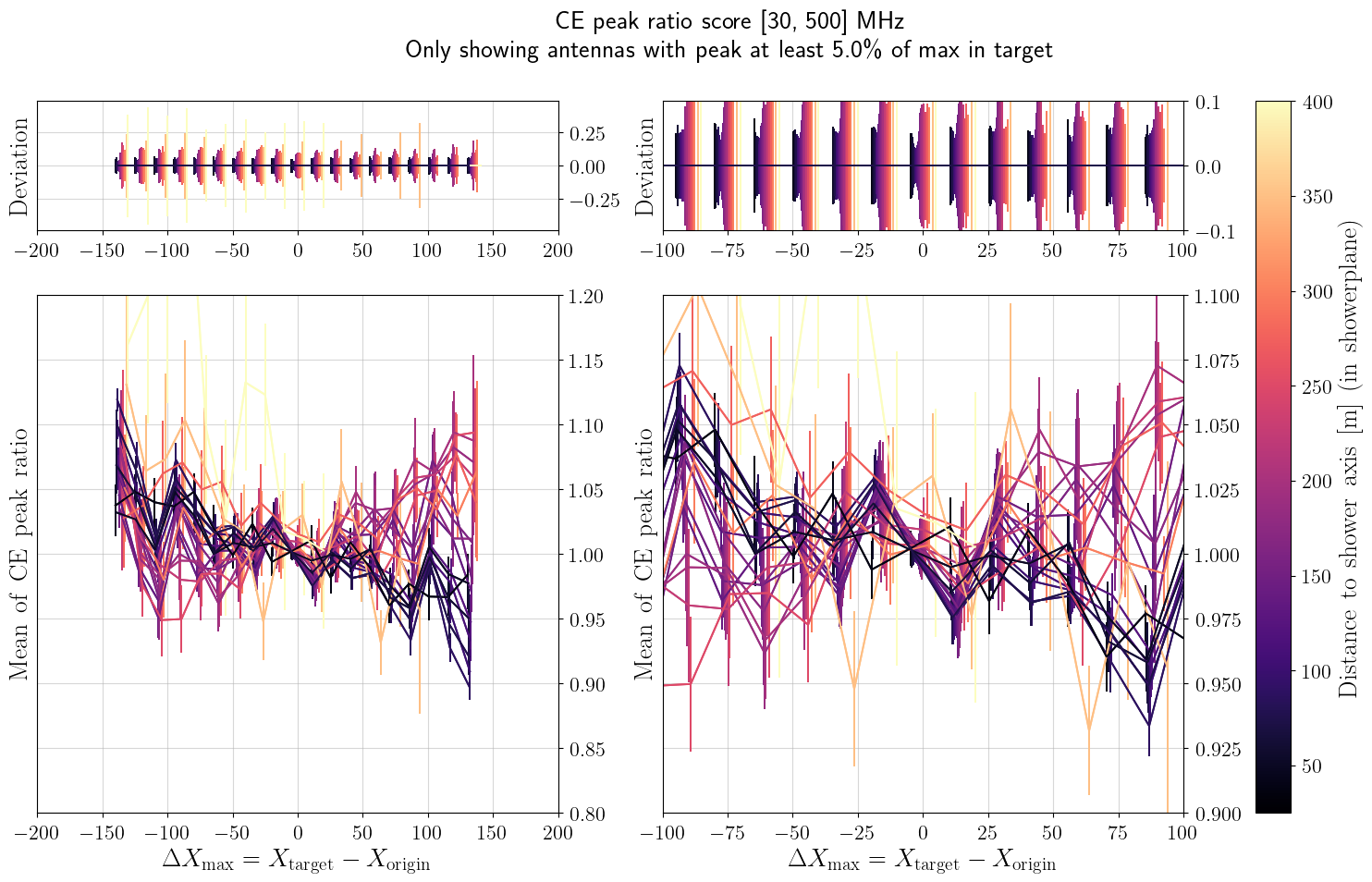}
    }
    \caption{
        The peak ratio scores for the simulation set using the AERA settings, from subsection \ref{subsec:aera_showers}.
        All showers have the same geometry, with a zenith angle of 40\textdegree\ and an azimuth angle of 0\textdegree\ in the CORSIKA coordinate system.
        The observation level of the simulations is set to 1564\textm , which is different from the 0\textm\ which was used for all the other simulations.
        We also used a different magnetic field, with values of $19.69 \; \mu \text{T}$ and $14.14 \; \mu \text{T}$ in the northern and upward direction, respectively, corresponding to the one typically used for simulations at the AERA site.
        The energies are still sampled from a log-uniform distribution between $10^{17} \eV$ to $10^{19} \eV$.
        In the GEO case (left) the antennas above $\sim 300 \m$ are removed because of the requirement that the peak amplitude is at least 5\% of the maximal amplitude in the footprint.
        This cut has less impact in the CE case (right) because the amplitudes in the antennas are much closer to each other.
        While this makes the standard deviations bigger, it has little effect on the mean values.
        They still follow the same trends as the GEO case, which are both in agreement with the general benchmarks.
    }
    \label{fig:benchmark_aera}
\end{figure*}

In Figure \ref{fig:benchmark_aera} we show the results of the benchmarking, which is carried out in the same way as detailed in the previous section.
As we can see, the mean values follow the same trend as our results from Section \ref{sec:benchmarking}.
The standard deviations are a bit larger, which are most probably due to the much smaller sample size.
In the case of the GEO emission, the antennas furthest from the shower axis are cut by our 5\% requirement which is why they do not appear in the plot.
The cut removes fewer antennas from the CE case, because the emission is much more similar in all antennas.
We suspect this is the reason for the increased scatter in this case.
Still, for the antennas in which we expect a sizeable CE contribution (between 100 and $300 \m$ from the shower axis) we do have standard deviations which are on the same level as the general benchmarks from the previous section.  

Given that within the uncertainties the results overlap with the general benchmarks, we conclude that our method correctly accounts for the magnetic field and observation level, and that the spectral functions we extracted previously do indeed generalise in this case.

\subsection{Using showers with high-density atmospheres} \label{subsec:change_atm}

Another important factor in the simulation is the type of atmosphere. 
Until now, we ran all our simulations with the US standard atmosphere as parametrised by Keilhauer.
Here we want to test if we can use our spectral functions with simulations that use a different atmosphere (while still keeping the atmosphere fixed during the synthesis process).
For this we created an artificial atmosphere by taking the one we had before and increasing the density by 10\%.\footnote{In CORSIKA, we achieved this by using ATMOD 0 and giving the layer parameters manually. These were the parameters from model 17, but the $a$ and $b$ parameters of the first four layers we multiplied by 1.1.}
This change is rather extreme, and corresponds to the largest difference in air pressure measured in The Netherlands last year.

Together with changing the atmospheric density profile, we also adapted the refractive index gradient.
We still use the Gladstone-Dale model, but the refractivity at sea level is also increased by 10\%.
Hence, the refractive index in this simulation set is $n_0 = 1.000321$.
This was calculated as $[1.1 \cdot (1.000292 - 1)] + 1$.
All simulated showers have the same geometry, which we set to 40\textdegree\ zenith angle and coming from the South.
The magnetic field was again the one from LOFAR.

While changing the mass overburden does not change the longitudinal profile of the showers, as these are expressed in \textgcm, it will change the geometrical proportions of it.
Also the geometrical size of our slices will change, since we fix their size to $5 \gcm$.
Therefore there will be many more of them.
However, we can still apply the same principles to synthesise the emission from showers with a different profile, given that we use the new density and refractive index profile when evaluating the spectral functions.

\begin{figure*}
    \centering
    \subfloat[Geomagnetic component]{
        \includegraphics[height=8.5cm, trim={17.1cm 0 3.5cm 2.5cm}, clip]{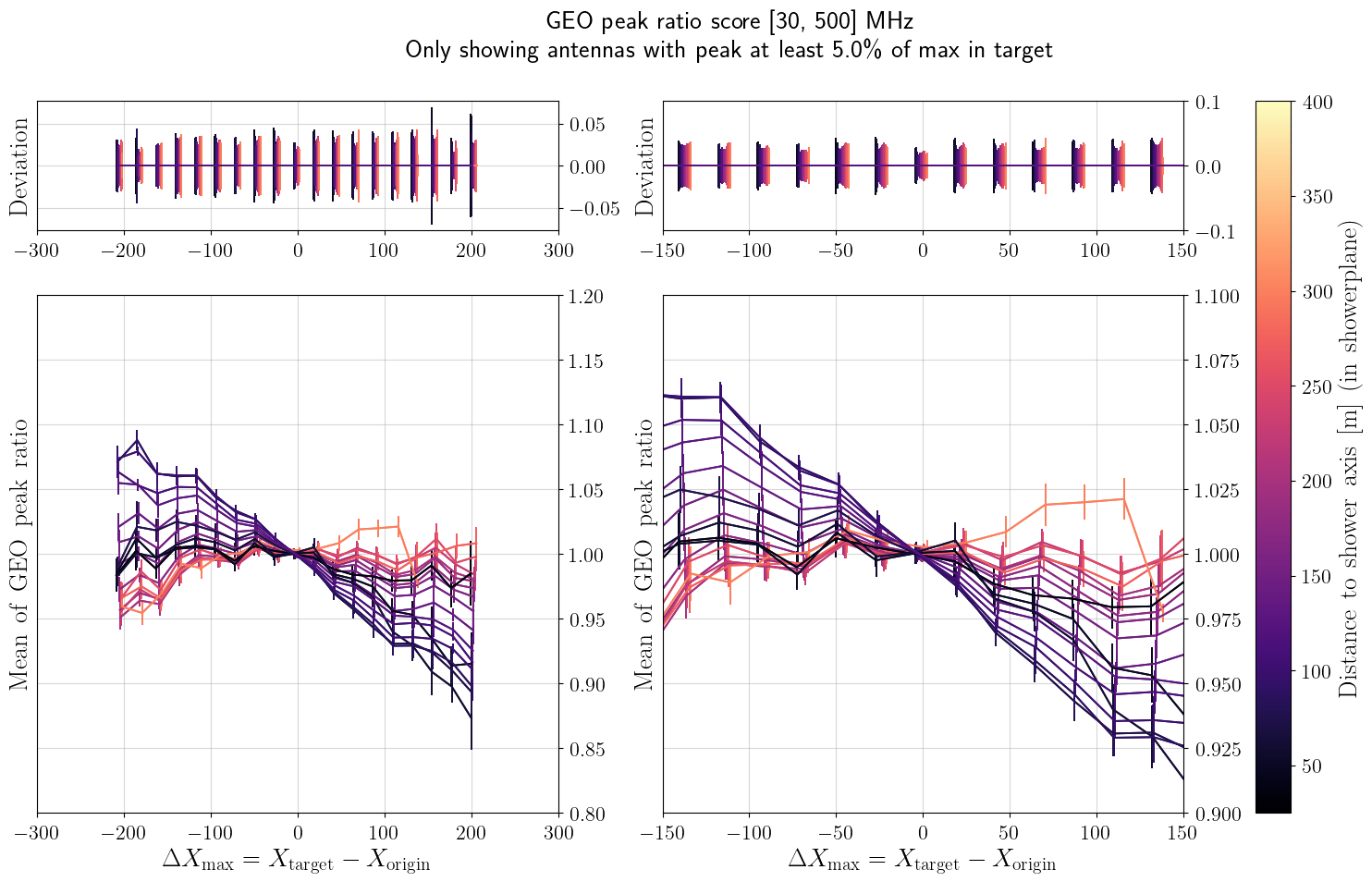}
    }~
    \subfloat[Charge-excess component]{
        \includegraphics[height=8.5cm, trim={17.1cm 0 0 2.5cm}, clip]{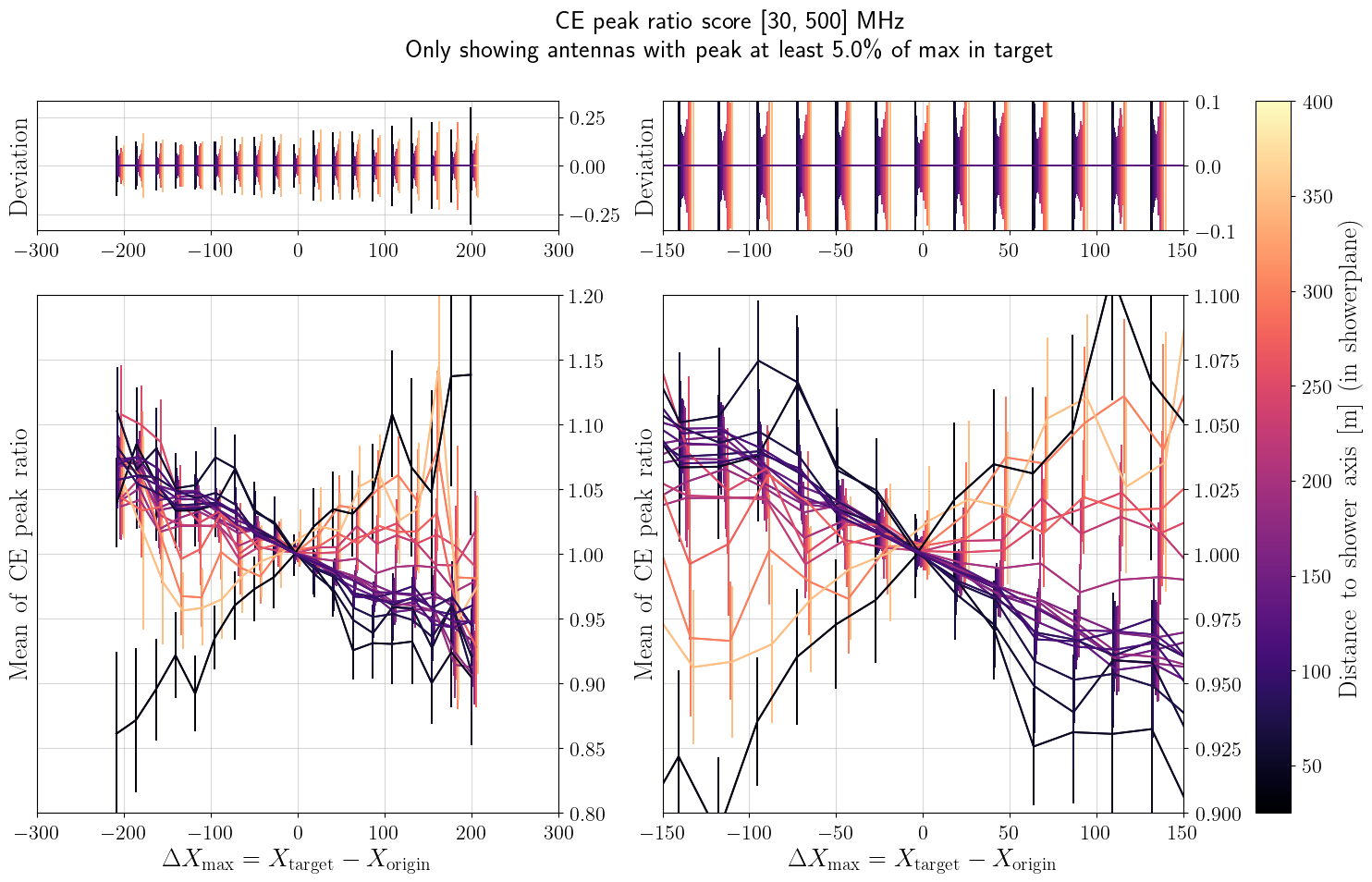}
    }
    \caption{
        The results of the peak ratio metric for the shower set which was simulated with an artificially heavy atmosphere, from subsection \ref{subsec:change_atm}.
        Here we used the LOFAR magnetic field and set the observation level back to 0\textm\ above sea level.
        The geometry of the showers is again 40\textdegree\ zenith angle, coming from the South.
        We did however use an artificial atmosphere, where we increased the mass overburden of the US standard atmosphere (by B. Keilhauer) by 10\%.
        Correspondingly, we increased the refractivity at sea level also by 10\%.
        We see again that the antennas above $\sim 350 \m$ from the shower axis are removed in the GEO plots (left) because of the cut on peak amplitude, which is not the case for the CE plots (right).
        This increases the standard deviations for the antennas very close to and very far away from the shower axis.
        In the antennas where we expect the strongest signal we do see a clear dip in the standard deviations, which for those antennas stay around 6\%.
        The trends of the mean values, as well as the standard deviations, are similar to those from the general benchmarks.
    }
    \label{fig:benchmark_atm}
\end{figure*}

The trends in Figure \ref{fig:benchmark_atm} are similar to what we observed in Figure \ref{fig:benchmark_aera}.
In the GEO case the antennas farthest from the shower axis are again cut away by the check on minimal peak amplitude.
The standard deviations are on the same level as the general benchmarks, between 3\% and 5\%.
For the CE we also see that fewer antennas are removed by the check.
This is very apparent by the fact that the antenna at $25 \m$ from the shower axis (the black line that runs from the lower left to the top right) is not removed.
We do not expect any significant signal in this antenna, so it should not really be considered.
If we only look at the antennas in the $100-300 \m$ range, we observe mean values and standard deviations which are similar to the general benchmarks.

Due to the increased mass overburden, the shower develops much farther away from ground.
The increased distance results in overall lower amplitudes.
This might explain why the cut on amplitude works even less well in CE case, as numerical errors might become relevant.

\subsection{Synthesising a shower with a different geometry than the template} \label{subsec:diff_geometry}

Lastly, we also want to explore the use case where we want to vary the geometry of the shower by up to a few degrees.
This use case is relevant for fitting the arrival direction of a cosmic ray event.
In the case of LOFAR, the particle detector array which triggers the read-out gives an estimate of the arrival direction with an accuracy of about 1\textdegree .
When processing the radio data, we would then generate a CoREAS simulation with that geometry \cite{Corstanje2021}.
During a reconstruction we can further refine the arrival direction estimate using the radio signal, and hence we would like to be able to quickly vary this direction without having to regenerate a new origin shower.

For this test, we simulated two base showers, one with a proton primary and one with an iron primary.
They both had a zenith angle of 30 degrees and an azimuth angle of 45 degrees.
Then, we simulated 18 additional showers, 9 for each primary.
Their arrival directions were varied with respect to the base case up to 8 degrees in zenith angle and 5 degrees in azimuth angle.
For the antennas we chose to always take the same positions in the shower plane.
Of course these project to different positions on the ground for each geometry.
When comparing the synthesis result to the CoREAS trace, we compare the result of synthesis in the antenna which has the same position in the shower plane to the simulation.

The benchmarking is again done in the same way as before.
However, in this case we split the results based on zenith angle.
Namely, when using a specific origin shower, we can only reasonably expect to be able to synthesise the emission from showers with a zenith angle which is lower or equal to that of the origin.
Otherwise the target profile would contain slices which are not present in the template.
For these we thus have no way of producing a synthesised signal.
Still, the algorithm implemented in the \texttt{SMIET} software will not crash when trying to synthesise a shower with higher zenith angle.
It will simply return the emission for all the slices the template does contain.
Since there is generally little emission in the last slices, it could still be reliable, although we did not look at this case in much detail.

\begin{figure*}
    \centering
    \subfloat[Geomagnetic component]{
        \includegraphics[height=8.5cm, trim={17.1cm 0 3.5cm 2.5cm}, clip]{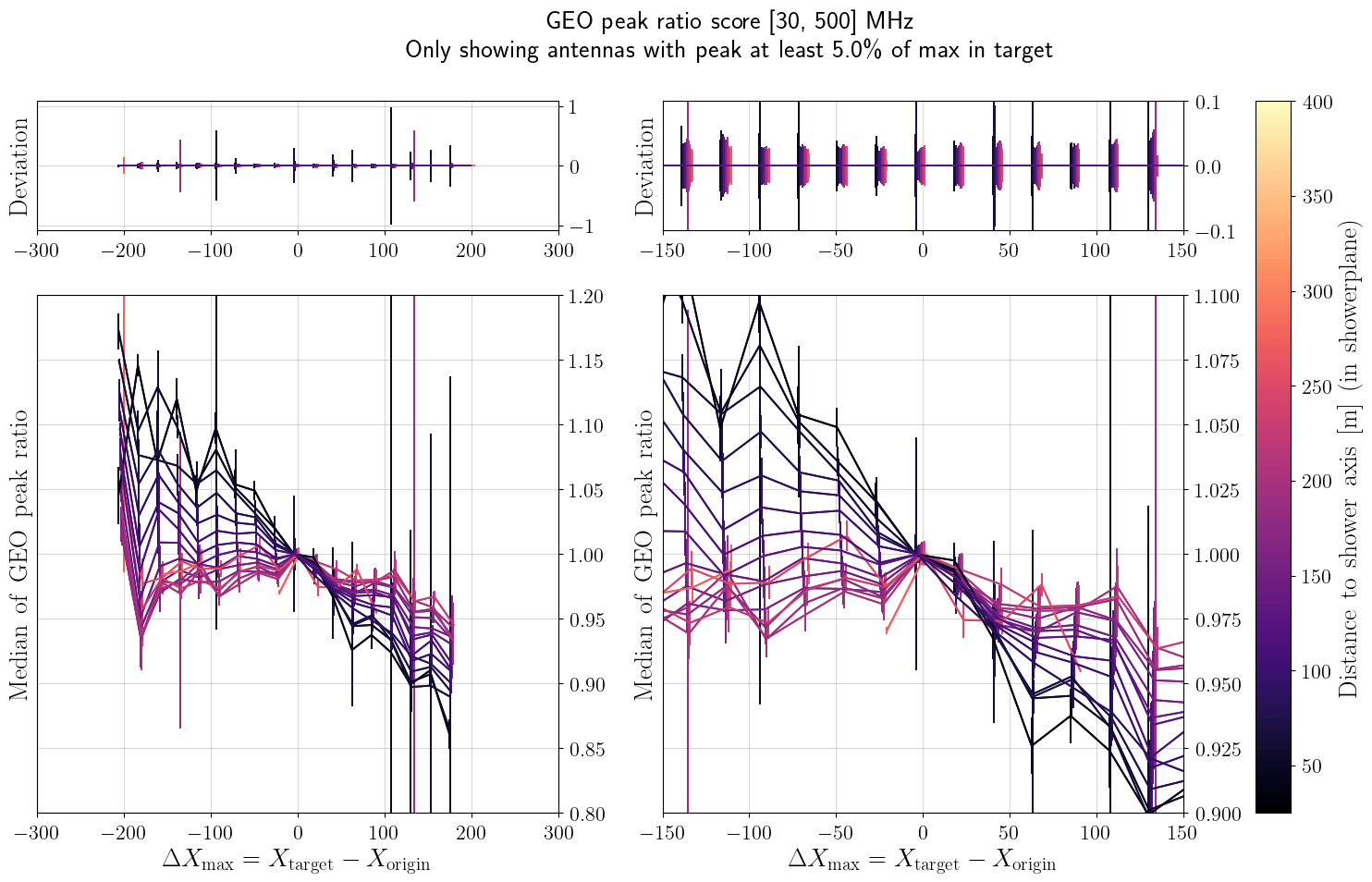}
    }~
    \subfloat[Charge-excess component]{
        \includegraphics[height=8.5cm, trim={17.1cm 0 0 2.5cm}, clip]{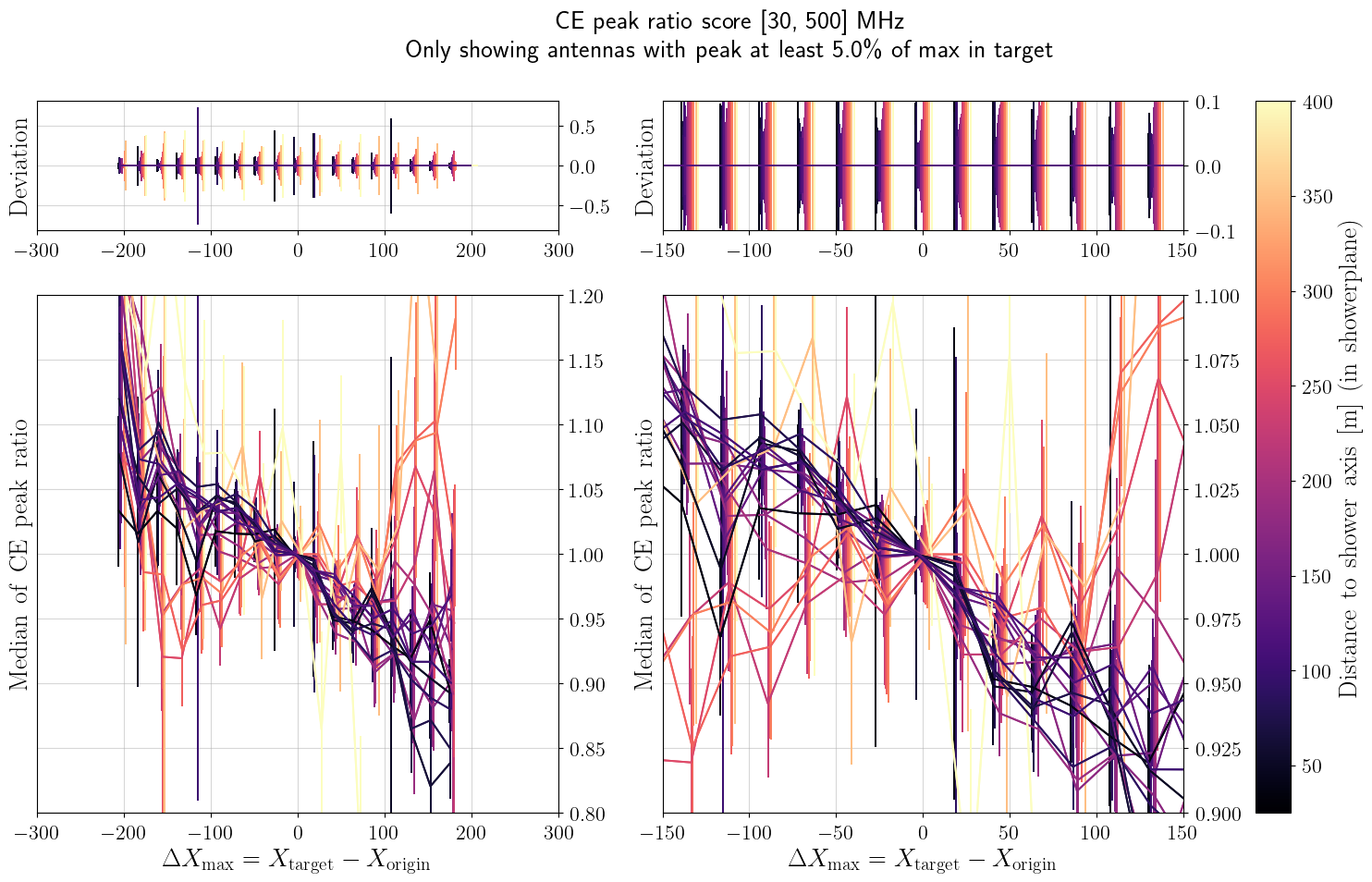}
    }
    \caption{
        The peak ratio scores for the test case where we vary the geometry of the showers (see subsection \ref{subsec:diff_geometry}).
        In this case the observation level and magnetic field were set at the same values as the general benchmarks.
        The energies were randomly sampled from the same distribution, a log-uniform disbtribution between $10^{17} \eV$ to $10^{19} \eV$.
        The showers all had different zenith and/or azimuth angles, except for two base showers (one initiated by proton and one by iron).
        The differences with respect to these base showers were up to 8 degrees in zenith angle and 5 degrees in azimuth angle.
        To make this plot, we selected the instances where the origin shower had a zenith angle equal or larger than the target shower.
        Note that we plot the median score instead of the mean, because of the outliers coming from the synthesis across the largest zenith angle difference.
    }
    \label{fig:benchmark_geometry}
\end{figure*}

As we can see in Figure \ref{fig:benchmark_geometry}, the benchmark follows again the same trends.
However, here we plot the median value instead of the mean.
The reason is that the mean scores are heavily influenced by outliers.
These outliers also result in the much increased standard deviations.

\begin{figure}
    \centering
    \includegraphics[width=\linewidth, trim={0 0 0 1cm}, clip]{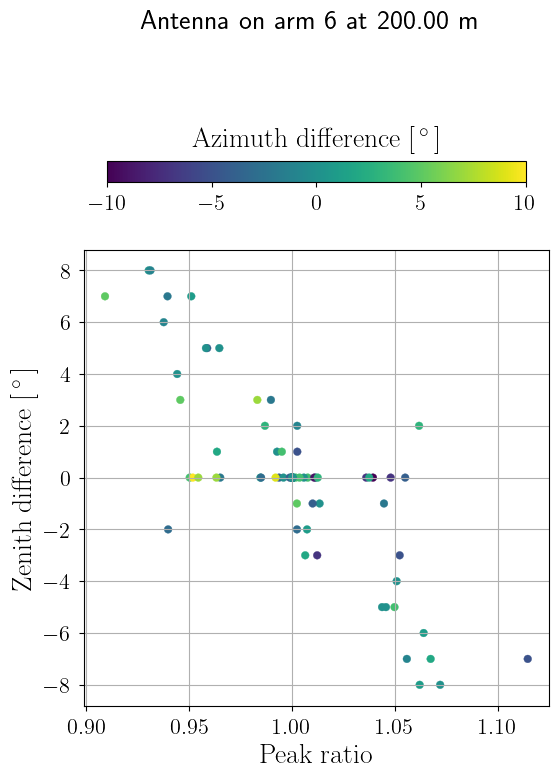}
    \caption{
        The peak ratio scores for individual synthesis cases where $| \DXmax | < 25 \gcm$, using showers from the set with different geometries.
        In this range we expect minimal biases from the template synthesis approach itself.
        Here we show the scores in an antenna on the $\vvB$ axis, at 200 \textm\ from the shower core.
        These are the entries that are used to calculate the median score in the central bin of Figure \ref{fig:benchmark_geometry}.
        We relate the score to the difference in zenith angle between target and origin shower.
        The points are coloured based on the difference in azimuth angle.
        We do not see any dependency on the azimuth angle, which is to be expected.
        The zenith angle does have a strong influence on the result though.
        The peak ratio score drops quickly as the difference between zenith angle of origin and target shower becomes larger.
        However, for any practical purposes, where changes in zenith angle would be on the order of one degree, the deviations are minimal.
    }
    \label{fig:benchmark_geometry_dependency}
\end{figure}

To understand where these outliers come from, we can look in more detail at how the peak scores depend on the difference in geometry in Figure \ref{fig:benchmark_geometry_dependency}.
We can see that they depend strongly on how much we vary the zenith angle.
A change in zenith angle will to first order change the arrival time of the signals.
We correct for this change using the phase corrections, as explained in subsection \ref{subsec:phases}.
While this does improve the scores (without the phase corrections the peak ratio scores ranged from 0.5 to 1.5), there are clearly some secondary effects at play.
We do note that these correlations between the peak ratio score and the zenith angle depend on the antenna distance to the shower axis.

Overall, we can conclude that it is possible to vary the geometry of the target shower with respect to the origin within a couple of degrees in zenith angle.
While we did not perform a statistical analysis over the entire antenna set, we can say that from manual inspection it appears that as long as the difference in zenith angle is not larger than $\sim 3^{\circ}$, the peak ratio score stays within one standard deviation.
Changes in azimuth do not introduce any changes in synthesis quality, as long as we properly rotate the antennas (which we achieve by ensuring the positions match in the shower plane).

\section{Discussion} \label{sec:discussion}

Here we wish to touch upon a few aspects which we have not discussed yet.
First of all, we want to highlight that template synthesis is not only fast, but also fully differentiable.
We explain in subsection \ref{subsec:differentiable} how we achieved this using a relatively new Python framework.
Because of this feature, template synthesis can be used in new analyses which use machine-learning techniques.
But thanks to its computational performance, the method can already be used in state-of-the-art analyses, where it will allow for reconstruction of more parameters than just \textXmax .
These avenues are explored in subsection \ref{subsec:nextgenanalysis}.
Then in subsection \ref{subsec:implications} we discuss some potential implications which arise from the observed scalings and universality in our approach.
Lastly we give an overview of potential improvements and future work in subsection \ref{subsec:future}.

\subsection{The differentiability of template synthesis} \label{subsec:differentiable}

From its inception, template synthesis was meant to be a fully differentiable forward model.
This means that it can readily be utilised in applications that require this property.
One obvious example would be machine-learning-based methods, which rely on back-propagation to train.

Another notable use case that we envision is using template synthesis for Information Field Theory (IFT) \cite{enslin_information_2019}.
This new and highly flexible reconstruction approach is based on Bayesian inference and likelihoods.
It fits the parameters of a physical model by comparing it to data samples in an iterative way.
The framework allows to specify prior knowledge about the parameters, which is taken into account during the fitting.
At the end, one recovers the statistical distribution of the optimised parameters.
This allows to not only find the best fitting values, but also determine the certainty on the model.
One crucial difference with machine learning is that it does not require a training phase.
Rather it infers all the parameters in ``one shot'' using all the data available.

In order for a model to be used in an IFT-based reconstruction, it needs to be fast and differentiable.
During the iterative procedure, the model is evaluated many times, so the speed with which a function call is made is key.
These requirements influenced the design of template synthesis in two key aspects.
Firstly, we explicitly separated the template generation process from the synthesis step.
This was not obvious, because due to the many divisions and multiplications it appeared more numerically stable to combine these.
However, thanks to this change it is possible to save templates to disk after they have been created.
Since this is by far the more expensive operation, it allows for much faster execution when sampling many different longitudinal profiles.

Secondly, we rewrote the entire codebase in JAX \cite{james_bradbury_jax_2018}.
JAX is a numerical computation framework akin to NumPy, but offering features like just-in-time compilation and GPU/TPU acceleration.
It also has built-in differentiation transformations, which make it easy to compute gradients.
Originally intended for machine learning research, the JAX framework has been adopted by the IFT community as its de facto framework.
Both the NumPy and the JAX implementations of the template synthesis code are available in the same repository.
Care has been taken to make migrating a script from one to the other as seamless as possible.
We do however note that JAX obfuscates many of the computations and is therefore more difficult to interpret.
New users are encouraged to use the NumPy implementation first and only move to the JAX version if they need raw performance.

\subsection{Next-generation analyses with template synthesis} \label{subsec:nextgenanalysis}

The ability to generate hundreds or thousands of shower for any observation in a short amount of time will open up new avenues for analyses.
In the previous section we already mentioned the IFT approaches, which could enable us to reconstruct the entire longitudinal profile. \cite{keito-ift-2025}
But even using a more classic $\chi^2$-fitting routine we can reconstruct more than just the shower maximum.
In \cite{corstanje_prospects_2023} the authors found a linear combination of $L$ and $R$, the shower width and asymmetry respectively, that could be reconstructed from densely sampled radio-emission data given enough simulations.
The latter part is prohibitive, as generating the necessary amount of simulations for each event would require enormous amounts of computational resources and time.
However, with template synthesis we can keep the computational impact similar to the standard \textXmax\ reconstruction.
Furthermore, since we can control the parameters of the longitudinal profile very precisely, we surmise that we could reconstruct both the width as well as the asymmetry separately.

Another application that we think would be fruitful to explore is to connect template synthesis to interferometric techniques. 
Previous work on deconvolving the detector effects and shower development used purely a macroscopic approach \cite{scholten_aperture_2024}.
Coupling this to template synthesis could offer new insights.

\subsection{Implications for air shower physics} \label{subsec:implications}

It was reported before that proton and iron showers gave different signatures in their radio emission \cite{carvalho_determination_2019}.
This is of interest to us, because template synthesis uses the principle that the primary particle type has no influence on the radio emission.
All the information comes from the longitudinal profile of the electromagnetic cascade.
In all of our testing, we could not find any discernable difference between the two particle types.
To investigate deeper we split the benchmarking results from Section \ref{sec:benchmarking} per primary type.

\begin{figure*}
    \centering
    \subfloat[Geomagnetic component]{
        \includegraphics[height=8.5cm, trim={17.1cm 0 3.5cm 2.5cm}, clip]{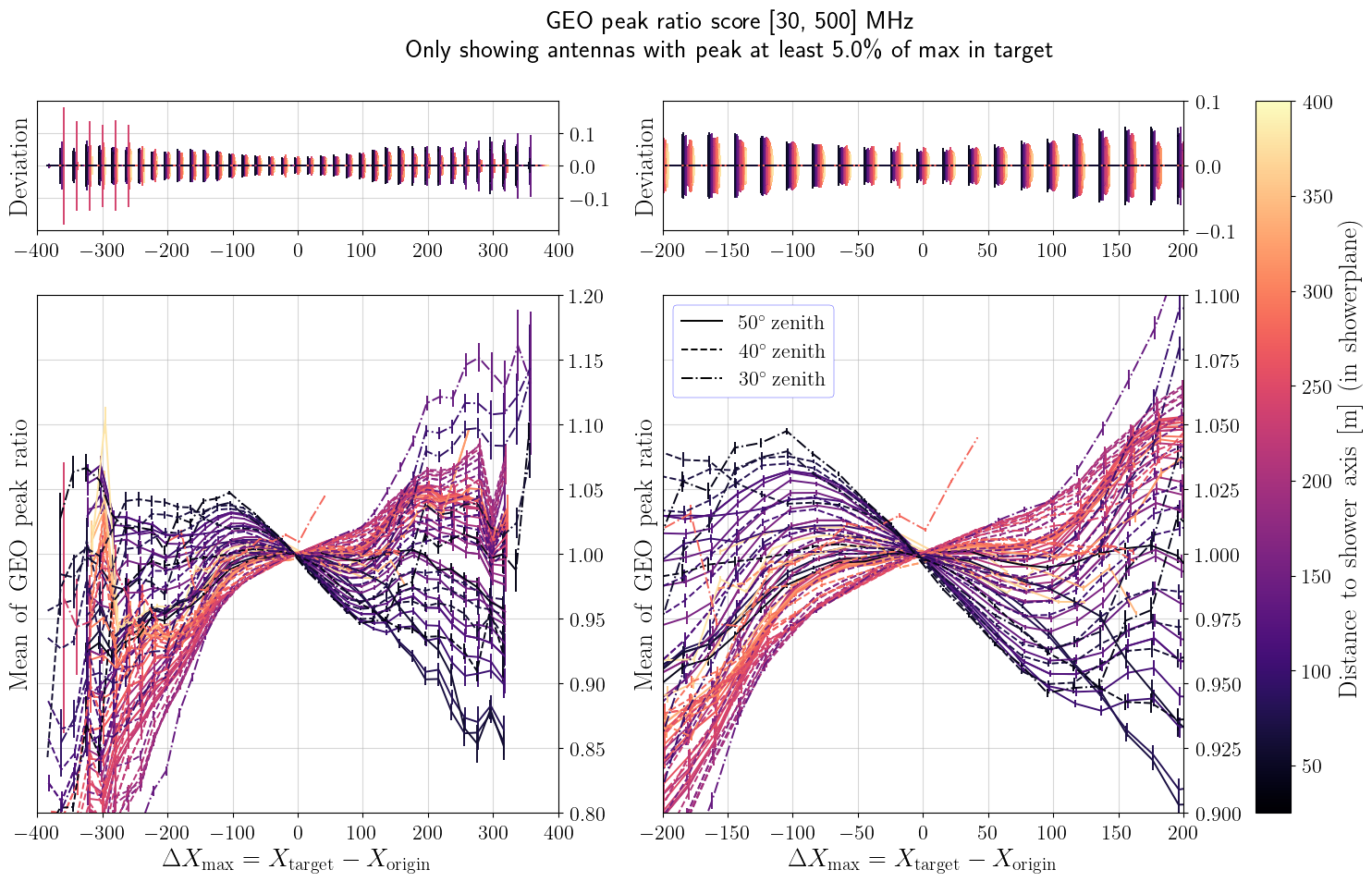}
    }~
    \subfloat[Charge-excess component]{
        \includegraphics[height=8.5cm, trim={17.1cm 0 0 2.5cm}, clip]{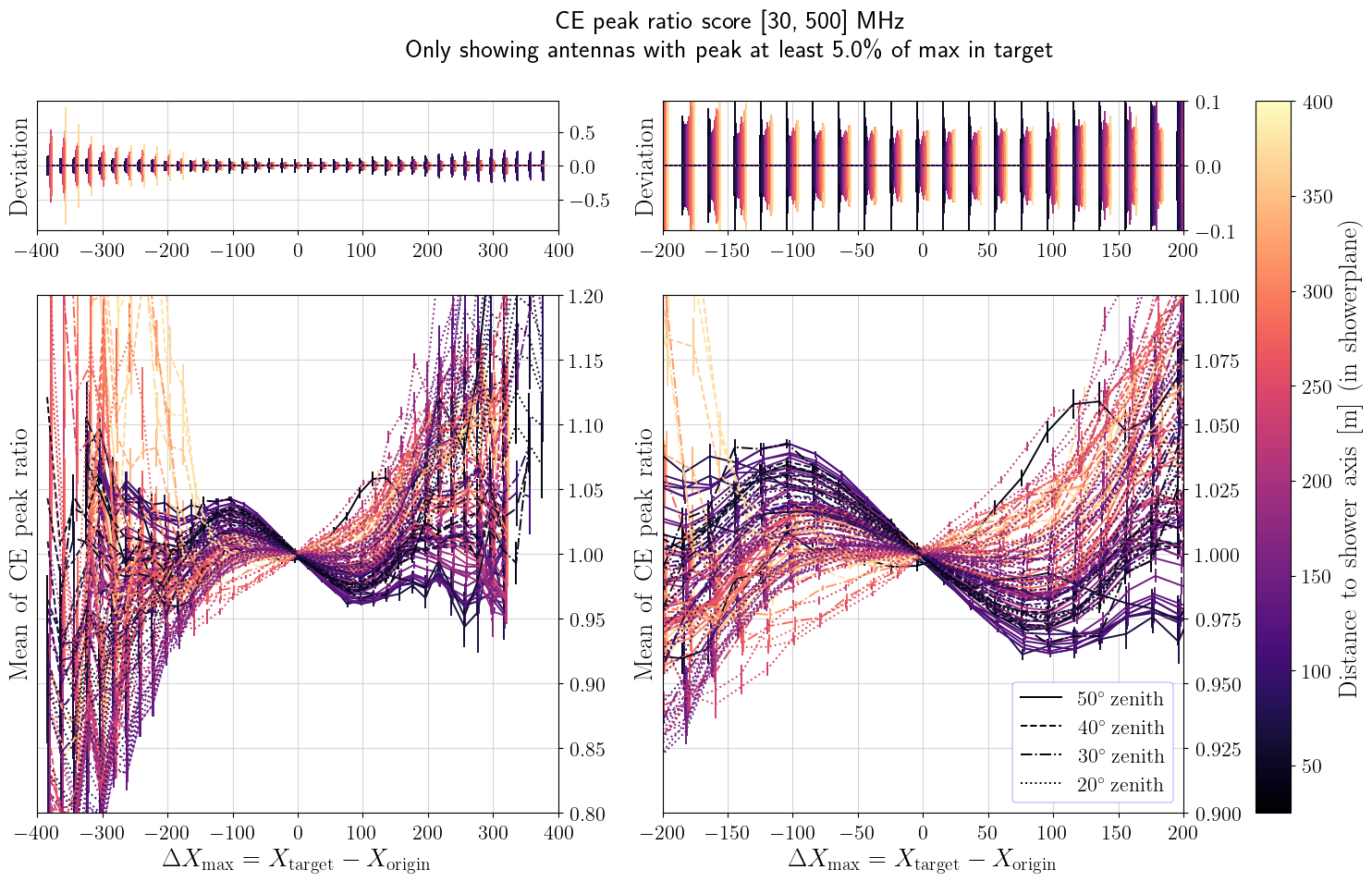}
    }
    \caption{
        The results of the benchmarks from Section \ref{sec:benchmarking}, but selecting only synthesis cases where the primary type of the origin and target are equal.
        The line that cuts off around $\DXmax \approx 50 \gcm$ in the GEO plot is the result of the 5\% cut in amplitude.
        This line corresponds to an antenna from the 30\textdegree\ zenith angle shower set, which sits at more than 300 \textm\ from the shower axis.
        When \textDXmax\ grows more positive the target moves closer to the ground and as a result this antenna falls outside of the footprint at some point. 
        In the region around $\DXmax = 0 \gcm$ there is practically no change to the results.
        Further out the mean values do shift, but the standard deviations are also larger.
        This is due to the fact that these bins are mostly populated by synthesis coming from mapping proton showers to iron showers or vice versa, as the mean \textXmax\ values of both elements are different.
        From this we cannot see any difference in the radio emission from these two primaries, apart from the differences in the longitudinal profile.
    }
    \label{fig:same_primaries}
\end{figure*}

In Figure \ref{fig:same_primaries} we show the subset of peak ratio scores which come from synthesising a target shower only with origin showers of the same primary particle type.
While the mean peak ratio scores do shift slightly, the changes are mostly in the bins with high \textDXmax .
These bins probably contain fewer entries in this case, since the \textXmax\ distributions of proton and iron showers peak at different values.
This is also visible in the uncertainties which are bigger.
However, for $\DXmax \in [-100, 100] \gcm$ there is practically no change in the mean scores.
Hence we conclude that we do not observe any statistically significant difference between the two cases.

We also want to come back on the use of \textDXmax\ in our method.
We made a similar analysis as in Figure \ref{fig:spectra}, but selecting slices where the showers had the same shower age
\begin{align*}
    s &= \frac{3t}{t + 2t_{\text{max}}}, \\
    \text{where} \quad t &= \frac{X}{X_0} \; ,
\end{align*}
instead of the same \textDXmax .
While the spectra also overlapped after rescaling, the remaining scatter in the spectra was on the same level and by eye even looked slightly larger than the case we discussed here.
This is an interesting observation because most shower properties like the energy distribution are usually universally described in terms of shower age.
Moreover the mapping from \textDXmax\ to shower age is non-linear.
In \cite{Lipari2009} however, the author presents an expansion of the shower age into a quantity that is a multiple of \textDXmax .
In that paper they discuss a different definition of shower age which relates to the slope of the longitudinal profile.
More recently in \cite{mendizabal_revisiting_2025} the authors also offer an improved longitudinal description of electromagnetic air showers, which results in a different relationship between shower age and atmospheric depth.
This could be related to what we see here, in the sense that one of the defining factors of the emission from a slice might be rate of change in the particle number in that slice.
That variable would for example definitely influence the current variation which is responsible for the geomagnetic component. 

\subsection{Further prospects} \label{subsec:future}

The template synthesis method is now ready to be used by the astroparticle community.
Still, there are a couple of aspects that could be investigated deeper in future work.
The first of which is related to what we discussed in the previous section.
Our suspicion is that the emission pattern of a slice is strongly (cor)related to the energy distribution of the electrons and positrons in that slice.
These distributions can be related to the age of the shower in the slice, such that they become highly universal \cite{Giller2005}.
In turn this could explain why the synthesis process works from a physical point of view.
Using a modified version of CoREAS, one can extract the energy distributions of every slice.
We believe it would be enlightening to analyse these and perhaps see if they follow similar trends as the spectral parameters.

Then there are also some improvements which can be made to the method in the future, as the standard deviations from the benchmarks were slightly larger than the scatter expected from the shower-to-shower fluctuations. 
As we mentioned, we relied on interpolation to get the signal at the viewing angles we desired in every slice.
However, it is possible to simulate the antennas in CoREAS at the exact viewing angles, since one can configure the antennas to have different positions for each slice.
We implemented the functionality to generate the input files for this case in the \texttt{SMIET} package.
The reason we did not perform this analysis is because simulating the required amount of showers to extract the spectral parameters is computationally expensive, and we already achieve good performance with the current version.
But if refinements to the method were desired in the future, this could be a good place to start.

Another possible future addition is to study and parametrise the changes in the phase spectra due to viewing angle, which may improve the range over which the zenith angle can change from input to target shower.

One more improvement which would benefit template synthesis a lot, is a better decoupling scheme for the two emission components.
Currently this is based on the position of the antenna in the shower plane.
On the $\vB$ axis, this completely fails.
Models to better isolate the GEO component exist for highly inclined air showers, which if generalised could also benefit our method \cite{schluter_signal_2023}.

Lastly we also want to touch upon the question up to which zenith angle template synthesis stays valid.
This will probably depend on whether the scaling relations we use in Section \ref{subsec:scaling_relations} remain valid for very inclined air showers.
In \cite{chiche_loss_2024} the authors report that from 70\textdegree\ onwards the GEO component scales differently with the air density.
This is similar to the results obtained in \cite{ammerman-yebra_density_2023} where the authors observed a loss of coherency if the density became very low.
It remains to be seen if this has a significant impact on template synthesis.
As we work per slice, the deviations from the scalings might be similar for both origin and target, effectively cancelling and resulting in a good synthesis result.
But we have not tested this so far.

In any case, from about 60\textdegree\ onwards, the Radio Morphing approach becomes valid.
As we mentioned before, we view template synthesis as a complementary method to this one.
Hence even if template synthesis would break down at high zenith angle, one could still combine the two to have an approach that can cover any zenith angle.

In Radio Morphing all emission is considered to originate from \textXmax .
This is like applying template synthesis with only one slice, which encapsulates the entire shower and has a ``position'' at the \textXmax\ of the shower.
So we expect that in the zenith angle range from 60\textdegree\ to 70\textdegree , both methods should converge.
An interesting analysis would be to check this explicitly, by comparing the results as well as by looking at the scaling relations both methods use and relate them to each other.

\section{Conclusion and future work} \label{sec:conclusion}

In this work we presented the generalised version of the template synthesis approach, of which we gave a proof of concept in \cite{desmet_proof_2024}.
The method combines elements from both microscopic and macroscopic approaches to the radio emission from extensive air showers.
Just like in other macroscopic approaches, we consider the emission to consist of two components: the geomagnetic and charge-excess ones.
The former results from the transverse current in the shower front, which varies as the shower propagates through the atmosphere.
The latter on the other hand is the consequence of a charge imbalance between the shower front, which carries with it a lot of electrons which are stripped off air molecules, and the positive ions that are left behind from said stripping.

Template synthesis also incorporates information from microscopic, Monte-Carlo approaches though.
It uses semi-analytical parametrisations of the amplitude frequency spectra which are extracted from big sets of CoREAS showers, which we call spectral functions.
We normalise these spectra according to a set of scaling relations, which we found correct for geometrical, atmospheric and magnetic effects.
These spectral functions then capture the behaviour of the amplitude spectra on the shower age, which is a measure of the stage of development the shower is in.

The spectra do show some fluctuations on top of the parametrisations that we introduced, however.
These are commonly referred to as shower-to-shower fluctuations.
To capture these in template synthesis, we start from a single microscopically simulated shower as an input.
We can then use the aforementioned parametrisations to rescale the emission based on the shower age.

Central to the whole procedure is the idea of atmospheric slicing. 
In every antenna the radiation is split into contributions from slices of a fixed atmospheric depth.
The total signal in an antenna can easily be retrieved by summing up all contributions.
But during the synthesis process we can use the fact that each slice acts as a point source to rescale emission based on the shower properties.
These rescaling are computationally easy, which makes the whole procedure feasible in a matter of minutes.
This is order of magnitude faster than the typical runtimes of Monte-Carlo based techniques.

In order to arrive at a description of our method which can be applied to all geometries with zenith angles up to 50\textdegree , we presented a geometrical setup which relates the antennas to the slices using their respective viewing angles.
Crucially, these were expressed as fractions of the Cherenkov angle in the slice.
This corrects for changes in the emission pattern when the refractive index in the slice changes, as the Cherenkov angle is dependent on that quantity.

We also introduced an explicit treatment of the phase frequency spectra.
In the template synthesis procedure we account for the arrival time of a signal in an antenna, by calculating the travel time from the slice as well as its emission time.
This has a significant effect on the coherence of the pulses coming from the different slices, which is a crucial factor when trying to synthesise a shower with a different zenith angle.

To verify the generality of template synthesis, we considered three different test scenarios.
First we used a simulation library which was generated with settings of the AERA site, which are different from the ones used to generate the spectral functions.
In a second scenario, we significantly changed the atmosphere used to simulate the showers.
The index of refraction was adapted accordingly.
In both cases we did not see a significant drop in synthesis performance.
We did notice, however, that the smaller amplitudes caused by the heavier atmosphere did play an important factor when benchmarking the method.

Lastly we also looked at synthesising a shower with a different geometry than the input shower.
This was possible thanks to the phase calculation we implemented.
We are able to synthesise showers up to 3\textdegree\ lower in zenith angle, with biases lower than 5\%.
Going to higher zenith angles will always remain a challenge, as in this case we have a section of atmosphere for which we have no template signal.
The azimuthal angle has no impact on the synthesis quality, as long as the antenna positions are rotated accordingly (which is taken into account by the viewing angle setup).

We benchmarked the method by comparing the results of template synthesis to traces coming from CoREAS.
Our main metric was the ratio of the peak in the synthesised trace to the peak in the CoREAS simulation.
From this we found a slight bias of synthesised traces, which seemed to depend on the distance to shower axis.
This bias was 0 when the difference in \textXmax\ between the origin and target shower was close to 0 \textgcm\ and grew as $|\DXmax|$ grew.
Interestingly the bias appears point-symmetrical around $\DXmax = 0 \gcm$, which makes it possible to correct for it by taking a weighted average of two synthesised traces, if we use one origin shower with an \textXmax\ higher than the target and a second origin shower which has a lower \textXmax . 

Similarly the deviations from the mean ratio also grew with $\DXmax$.
They were a few percent around $\DXmax \approx 0 \gcm$, which is on the same level as the shower-to-shower fluctuations we inherently expect.
When going to $|\DXmax| = 200 \gcm$ they grew up to 6\% and 9\% for the geomagnetic and charge-excess components respectively, which is well within the tolerances of most analyses and similar to results obtained for the fixed geometry case \cite{desmet_proof_2024}.
All of these results were obtained using the [30, 500] \textMHz\ frequency band.
When using smaller frequency ranges, these results improve.
In \cite{desmet_proof_2024} for example, we found that the standard deviations are halved when working in the smaller [30, 80] \textMHz\ range. 

In an effort to allow everyone to use template synthesis, we created a Python package which is now public.
We implemented the algorithm in two different frameworks.
The first one is NumPy, which most people are familiar with.
This version of the code is ideal for understanding the inner workings as well as debugging specific aspects, as it provides some convenience classes.
We also have a JAX implementation, which has the benefit of being much more performant as well as fully differentiable out of the box.
This makes it easy to slot the package into machine learning or information field theory applications.
An interface into the NuRadio framework is planned and will be released in the near future.

We are excited to see all the applications for which template synthesis will be used.
A first analysis we are working on is a LOFAR-style \textXmax\ reconstruction using template synthesis instead of CoREAS simulation.
Next to this, we are also working on an IFT based reconstruction of the longitudinal profile of an air shower using the radio emission.
We also envision the possibility of connecting our method to interferometric techniques.

\section*{Acknowledgements}

Mitja Desmet is supported by the Flemish Foundation for Scientific Research (FWO-AL991).
He acknowledges the creative mastermind of Yarno Merckx, who invented the term "Smiet" during their bachelor education together.

The authors also want to thank Felix Schl\"uter for the inspiration for the interpolated synthesis approach.

This research was supported by Deutsche Forschungsgemeinschaft (DFG) -Projektnummer 531213488.
The authors gratefully acknowledge the computing time provided on the high-performance computer HoreKa by the National High-Performance Computing Center at KIT (NHR@KIT). This center is jointly supported by the Federal Ministry of Education and Research and the Ministry of Science, Research and the Arts of Baden-W\"urttemberg, as part of the National High-Performance Computing (NHR) joint funding program. HoreKa is partly funded by the German Research Foundation.

\bibliographystyle{elsarticle-num} 
\bibliography{references, manual_entries}

\clearpage
\onecolumn
\appendix

\section{Derivation of the arrival time} \label{app:arrival_time}

To find the emission time we start from the equation giving the mass overburden as a function of height above sea level \cite{Heck2023},
\begin{equation} \label{eq:T_z}
    T(h) = a_i + b_i \exp \left( -\frac{h}{c_i} \right) \; .
\end{equation}
Here the subscript $i$ refers to the multiple layers that used to describe the entire atmosphere.
In CORSIKA there are 4 layers, and a 5th one which describes the top end of the atmosphere with a linear decrease instead of an exponential.
The units of $a_i$ and $b_i$ are \textgcm , where as $c_i$ has units of \textcm .

Now we want to find the distance along the shower axis from the core to some slant depth $X_{\slice}$.
Assuming a flat Earth geometry, which is a valid approximation if the zenith angle of the shower is smaller then $\sim 60^{\circ}$, the corresponding vertical mass overburden is
\begin{align*}
    T = X_{\slice} \cdot \cos (\theta) \; .
\end{align*}
If we substitute this in the left hand side of Equation \eqref{eq:T_z} and solve for $h$, we find that
\begin{align*}
    h = - c_i \exp \left( \frac{X_{\slice} \cdot \cos ( \theta )}{b_i} \right ) \; . 
\end{align*}
Finally we convert this to distance along the shower axis by using Pythagoras theorem,
\begin{equation}
    d \cdot \cos ( \theta ) = - c_i \exp \left( \frac{X_{\slice} \cdot \cos ( \theta )}{b_i} \right ) \; .
\end{equation}

The integration of the refractivity is a bit more involved.
For this we begin with the Gladstone-Dale model, which relates the change in refractivity $N$ to a change in density $\rho$,
\begin{equation} \label{eq:gladstone_dale}
    \frac{N(h)}{\rho(h)} = \frac{N_{\text{sea}}}{\rho_{\text{sea}}} \; .
\end{equation}
We can then use the fact that the density is the derivative of the mass overburden as given in Equation \ref{eq:T_z}, since the mass overburden is the integrated density up to some height,
\begin{equation}
    \rho (h) = -\frac{b_i}{c_i} \exp \left(- \frac{h}{c_i} \right) \; . 
\end{equation}
If we use this in Equation \eqref{eq:gladstone_dale} and integrate this expression from some height $h_{\ant}$ to another height $h_{\slice}$, we find that
\begin{align*}
    N_{\text{eff}} 
    = \int_{h_{\ant}}^{h_{\slice}} N(h') \, \text{d}h'
    = \frac{N_{\text{sea}}}{\rho_{\text{sea}}} \cdot b_i 
    \left[ 
        \exp \left( -\frac{h_{\ant}}{c_i} \right)  - 
        \exp \left( -\frac{h_{\slice}}{c_i} \right)
    \right] \; .
\end{align*}
For a flat geometry, we can simply use the vertical coordinate for $h_{\ant}$ and $h_{\slice}$.
But this integration is also valid for curved geometries, as long as we calculate the height $h$ correctly.
Here we recognise the difference between the mass overburden at the two heights, which gives us the relation mentioned in the main text:
\begin{equation}
    n_{\text{eff}} 
    = 1 + N_{\text{eff}} 
    = 1 + \frac{N_{\text{sea}}}{\rho_{\text{sea}}} \cdot [T(h_{\ant}) - T(h_{\slice})] \; .
\end{equation}

\newpage
\section{More benchmarking plots} \label{app:more_plots}

\begin{figure}[hb]
    \centering
    \subfloat[Geomagnetic component]{
        \includegraphics[height=8.5cm, trim={17.1cm 0 3.5cm 2.5cm}, clip]{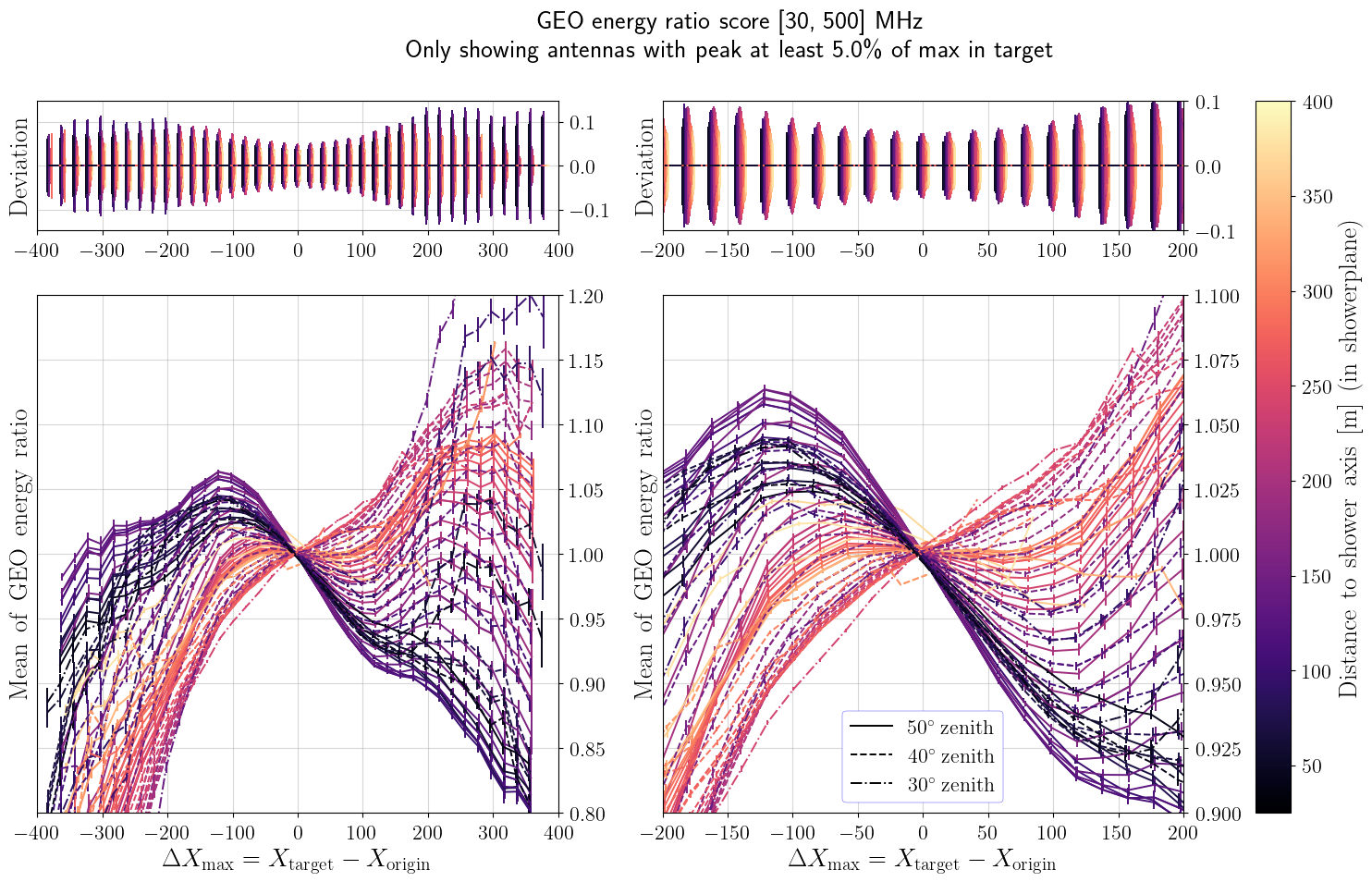}
    }~
    \subfloat[Charge-excess component]{
        \includegraphics[height=8.5cm, trim={17.1cm 0 0 2.5cm}, clip]{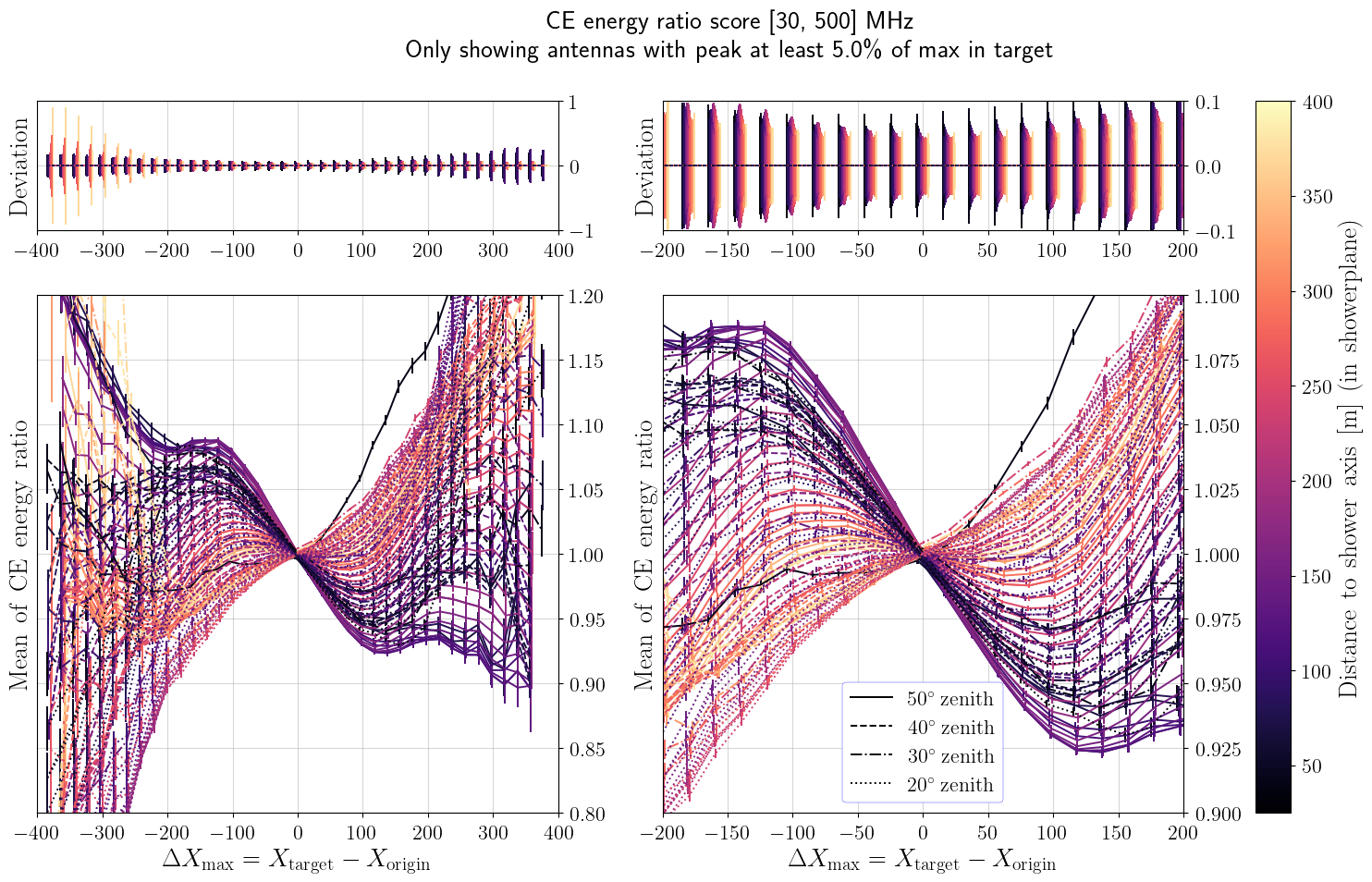}
    }
    \caption{
        The $S_{\text{fluence}}$ metric from Equation \eqref{eq:energy_ratio}.
        These results are consistent with the peak ratio results, accounting for the factor 2 that comes from the fluence calculation.
        Some antennas, notably those furthest from the shower axis in CE case, do not follow this pattern though.
        This probably hints at some remaining deviations in the pulse shape.
    }
\end{figure}

\begin{figure}[ht]
    \centering
    \subfloat[GEO]{
        \includegraphics[height=8.5cm, trim={17.1cm 0 0 2.5cm}, clip]{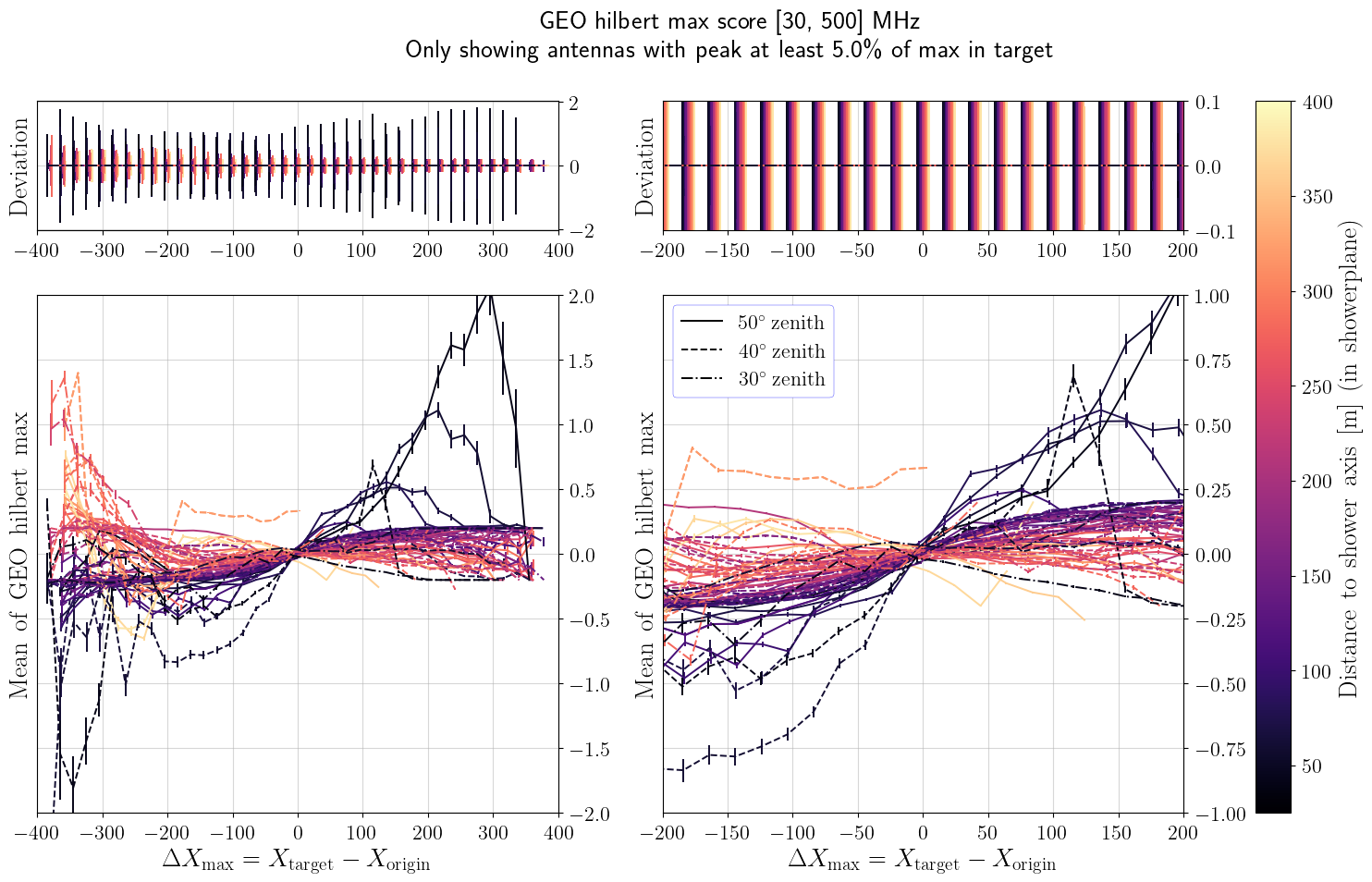}
    }~
    \subfloat[CE]{
        \includegraphics[height=8.5cm, trim={17.1cm 0 0 2.5cm}, clip]{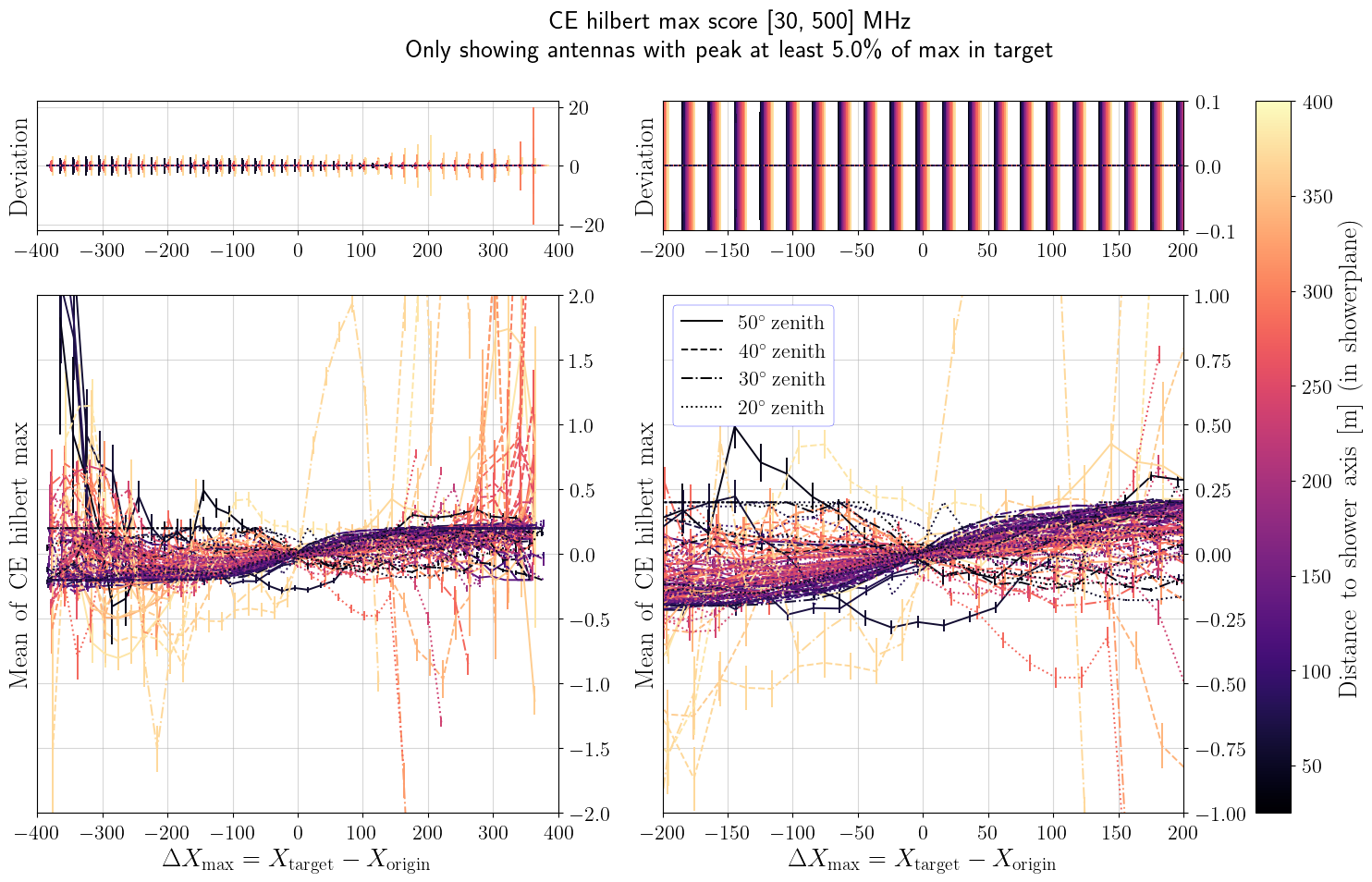}
    }
    \caption{
        The $S_{\text{Hilbert}}$ metric from Equation \eqref{eq:hilbert_max}.
        The sampling rate in all simulations was 0.2 ns , which is the unit on the y-axis.
        We can see that mean differences between the maxima of the Hilbert envelopes are always within one time sample.
        For weak signals the Hilbert envelope is not well-defined, which leads to some curves having very erratic behaviour.
    }
\end{figure}

\end{document}